\documentclass[%
superscriptaddress,
preprint,
 amsmath,amssymb,
 aps,
 pra,
floatfix,
showkeys
]{revtex4-2}

\usepackage{ulem}
\usepackage{graphicx}
\usepackage{subfigure}
\usepackage{xcolor}
\usepackage{dcolumn}
\usepackage{bm}
\usepackage{hyperref}

\begin{document}

\preprint{arXiv/v2}

\title{Three-state active lattice gas: a discrete Vicseklike model with excluded volume}

\author{Tiago Venzel Rosembach}
\email{tiagovenzel@cefetmg.br}
\affiliation{%
 Departamento de Física, ICEx, Universidade Federal de Minas Gerais, C. P. 702, 30123-970 Belo Horizonte, Minas Gerais, Brazil.
}
\affiliation{Departamento de Formação Geral de Leopoldina, Centro Federal de Ensino Tecnológico de Minas Gerais, Rua José Peres, 558 - Cento - Leopoldina-MG, 36700-001, Brazil.}
\author{Ana Luiza Novaes Dias}
\email{anand@ufmg.br}
\affiliation{%
 Departamento de Física, ICEx, Universidade Federal de Minas Gerais, C. P. 702, 30123-970 Belo Horizonte, Minas Gerais, Brazil.
}
\author{Ronald Dickman}
\email{dickman@fisica.ufmg.br}
\affiliation{%
 Departamento de Física, ICEx, Universidade Federal de Minas Gerais, C. P. 702, 30123-970 Belo Horizonte, Minas Gerais, Brazil.
}
\affiliation{Departamento de Física and National Institute of Science and Technology for Complex Systems, ICEx,Universidade Federal de Minas Gerais, C. P. 702, 30123-970 Belo Horizonte, Minas Gerais, Brazil.}

\date{\today}

\begin{abstract}
We study a discrete-space model of active matter with excluded volume.  Particles are restricted to the sites of a triangular lattice, and can assume one of three orientations. 
Varying the density and noise intensity, Monte Carlo simulations reveal a variety of spatial patterns. Ordered states occur in the form of condensed structures, which (away from the full occupancy limit) coexist with a low-density vapor. The condensed structures feature low particle mobility, particularly those
that wrap the system via the periodic boundaries.  As the noise intensity is increased, dense structures give way to a disordered phase.  
We characterize the parameter values associated with the condensed phases
and perform a detailed study of the order-disorder transition at (1) full occupation and (2) at a density of 0.1. In the former case, the model
possesses the same symmetry as the three-state Potts model and exhibits a continuous phase transition, as expected, with critical exponents consistent with
those of the associated Potts model.  In the low-density case, the transition is clearly {\it discontinuous}, with strong dependence of the final state upon the initial configuration, hysteresis, and nonmonotonic dependence of the Binder cumulant upon noise intensity.

\end{abstract}

\keywords{active matter; active lattice gas; phase transitions; critical exponents; correlation function; hysteresis}

\maketitle

\newpage

\tableofcontents

\section{\label{sec:int} Introduction}

Since the introduction of a model of self-organized motion of mobile agents by
Vicsek and coworkers nearly thirty years ago \cite{vicsek1995}, 
active matter (AM) has motivated enormous theoretical, simulational and experimental interest amongst investigators in statistical physics and allied fields.  Active matter, understood as collections of many interacting particles, each of which consumes free energy to self propel, is intrinsically far from equilibrium. 
Well studied examples of AM are groups of macro- or microorganisms
that interact to yield organized, collective motion. Subcellular processes exhibit diverse examples of AM. The cytoskeleton, for example, maintains polarization dynamics and stresses far from equilibrium via a chemical free energy supply; the plasma membrane, with ion pumps and actin polarization centers, behaves like an active fluid  \cite{ramaswamy2001, kumar2014, brugues2014, tatiane2017, joanny2013, prost2015, ramaswamy2000}.

The Vicsek model (VM), introduced in 1995 uses simple rules to describe the continuous-space dynamics of active particles aligning their direction of motion with that of their neighbors, leading to a phase transition between collective motion at high density and low noise and disordered motion in the opposite limit \cite{vicsek1995}. Subsequent studies showed that the VM and related models exhibit a regime marked by coexistence of a condensed, ordered phase and a disordered vapor \cite{solon2015c}.  The latter falls between a uniformly ordered phase and a disordered gas phase \cite{chate2020}. The VM represents {\it dilute} active matter in the sense
that excluded-volume interactions between particles are ignored.

The question naturally arises whether such coexistence occurs in 
discrete-space AM systems; it is in fact found in the active Ising model (AIM) \cite{solon2013,solon2015b} as well as in diverse lattice models. 
Similar to the (continuous-space) Vicsek model, the homogeneously ordered and disordered phases are separated, in the density/noise-intensity plane, by a
coexistence region.
Recent works investigating a generalization of the AIM, the $q$-state active Potts model (APM), show that as long as there is no restriction on the number of particles that may occupy a lattice site, phase coexistence is observed \cite{chatterjee2020, mangeat2020}. Another discrete-space active matter model is the $q$-state active clock model (ACM). Similar to what is observed in the VM and APM, the active clock model displays a liquid-vapor type transition. The ACM coexistence region features macroscopic phase separation for small $q$ values and microscopic separation for large $q$ values, as in the VM \cite{chatterjee2022, solon2022}.

Since a significant fraction of the available space can be occupied by the organisms constituting a bird flock or school of fish, it is important to understand how organization in AM is affected by excluded-volume (EV) interactions.
In AM models with EV, particle mobility decreases with density, leading to immobile structures such as traffic jams.
Congestion is observed, for example, in vehicular traffic, embryogenesis, tumor formation, and herds \cite{orosz2010,helbing2001,khakalo2018,moussaid2012}. Self-propelled motion of entities with EV can lead to structures similar to those found in the Vicesk and AIM, as well as traffic jams, in addition to ``mixed" configurations containing two or more simple structures \cite{chate2008a, chate2008b, peruani2011, karmakar2022}.

Previous studies of AM models with excluded volume include the four-state active Potts model proposed by Peruani \textit{et al}. \cite{peruani2011}, with four particle orientations on a square lattice. In this model, a parameter \textit{g} determines the intensity of alignment between particles occupying nearest-neighbor sites; as 
\textit{g} is increased at fixed particle density, the system exhibits three phases: disordered aggregates for weak alignment, a phase with local ordering characterized by traffic jams and gliders (dynamic traffic jams with two opposing orientations), and immobile bands that emerge under strong alignment. 

The present study is motivated by our interest in characterizing the
phases and associated phase transitions in a simple active-matter model
with EV and a minimal set of particle motions in the plane. 
Since EV is most efficiently treated in discrete space, and in the interest of simplicity, we study a lattice gas in which particle velocities 
are limited to three directions on the triangle lattice.  Our model includes a Vicseklike alignment interaction favoring the formation of ordered groups; this tendency, which follows a majority-vote scheme, is nevertheless frustrated by the reduced particle mobility in dense regions, leading to the emergence of condensed structures which may be characterized as bands or ``traffic jams".

Using numerical simulation we find that the model exhibits a variety of stationary states depending on the parameters (density, noise intensity, system size) and the initial configuration (IC). We find that the most stable configuration at low noise intensity is an immobile band of particles that wraps the system via the periodic boundaries, that is, it forms a structure that closes via the periodic boundaries. 
At high noise intensities, the steady state is disordered, having 
equal average particle fractions in the three directions, and
spatially uniform.
The model exhibits one or more phase transitions as one increases the noise intensity at fixed density. The order-disorder transition is discontinuous over most of the range of densities. In the limiting case of full occupancy, all particles are immobile, and the model possesses the permutational symmetry of the 3-State Potts model. We find that, as expected,
the phase transition belongs to the 3-state Potts universality class in this limit. 

The remainder of the article is organized as follows: In the following section we define the model and contrast it with previously studied discrete-space AM models.  In Sec.~\ref{sec:result} we report and discuss the results obtained via Monte Carlo simulation. Finally, Sec.~\ref{sec:concl} contains a summary of our main conclusions and prospects for future work.

\section{\label{sec:model} Model}
\begin{figure}[!htp]
    \centering
    \includegraphics[width=0.85\linewidth]{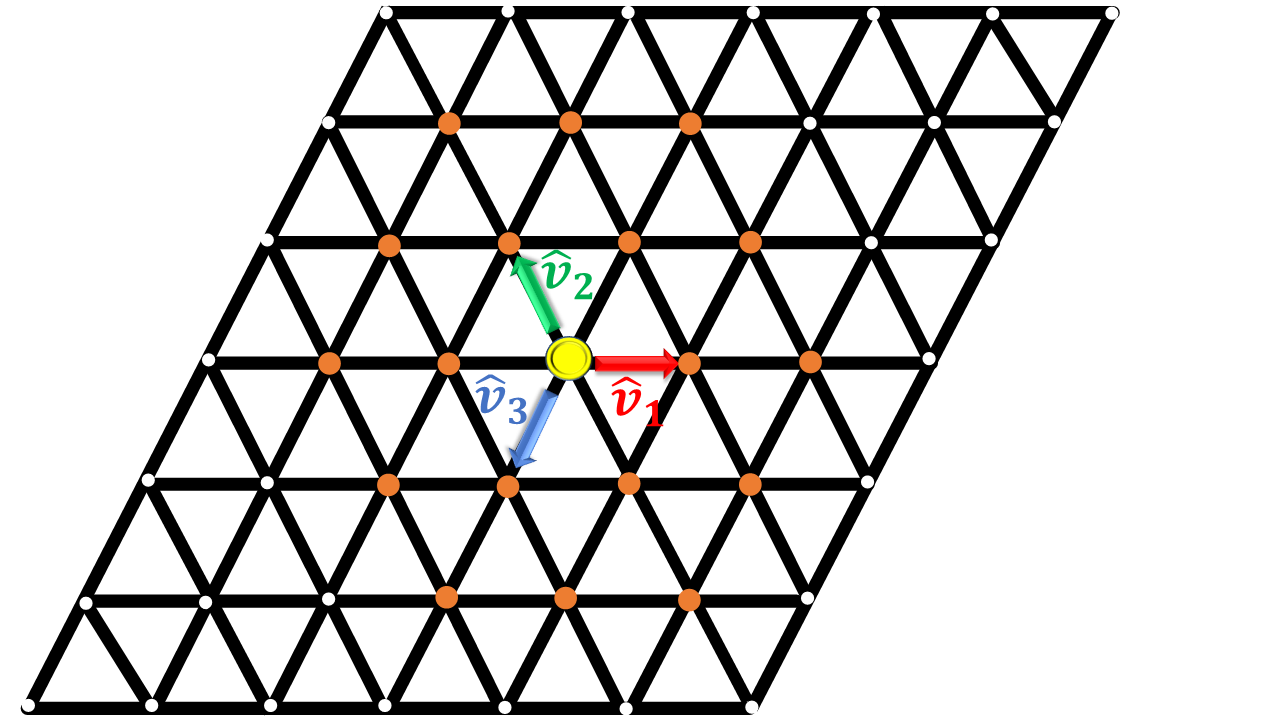}
    \caption{\label{fig:latitce} (color online) Schematic of the three-state Active Potts Model.  Sites in orange correspond to the nearest, second- and third-neighbors of the site marked in yellow.  The set of 19 sites comprising the central (yellow) site and its neighbors (orange) constitute the {\it neighborhood} of the yellow site.  Red, green and blue arrows correspond to velocities $\hat{v}_1$,  $\hat{v}_2$ and $\hat{v}_3$, respectively.  The system is periodic in both directions, so that all sites possess the same number of neighbors.}
\end{figure}

We consider a set of $N$ particles moving on a triangular lattice of $L^2$ sites with periodic boundaries. Volume exclusion is imposed via the condition that at most one particle may occupy a given site. Although each site has $z = 6$ nearest neighbors, velocities are restricted to a set of only three unit vectors: $\hat{v}_1 = {\bf i}$, $\hat{v}_2 = -{\bf i}/2 + \sqrt{3}{\bf j}/2$ and $\hat{v}_3 = -{\bf i}/2 - \sqrt{3}{\bf j}/2$.  This is the smallest unbiased set sufficient for a particle to travel from the origin to any other site in at most ${\cal O}(L)$ steps. Figure~\ref{fig:latitce} shows the three directions of allowed motion. 

The update rules for particle positions and velocities parallel those of the VM \cite{vicsek1995} in the context of discrete position and orientation.  At each {\it elementary event}, a randomly chosen particle, $i$, with current velocity ${\bf v}_i$, updates its velocity to ${\bf v}'_i$.  If the neighboring site in the direction of ${\bf v}'_i$ is vacant, the particle moves to this site.  Otherwise, the particle remains at its current position while maintaining ${\bf v}'_i$ as its velocity. We associate a time interval of $1/N$ with each elementary event.

The updated velocity of particle $i$ depends on the velocities of all particles in its {\it neighborhood}, which we take as the set of first, second, and third neighbors, as well as particle $i$ itself, for a total of nineteen sites. (Thus the neighborhood of a particle can never be empty.) The use of an extended interaction neighborhood improves statistics in simulations and reduces the likelihood of situations lacking a clearly defined majority orientation.
To update its velocity, particle $i$ performs a ``census'' of the velocities of all particles in its neighborhood. Each appraisal of a velocity is subject to error: with probability $\eta$, a particle with velocity ${\bf v}_j$ is 
{\it incorrectly identified} as having one of the other two velocities; this applies to particle $i$ as well.  The updated velocity ${\bf v}'_i$ is taken as the majority of the set of perceived velocities. In case of a two-way tie between velocities $\hat{v}_k$ and $\hat{v}_m$, one of the two is chosen at random for ${\bf v}'_i$; in case of a three-way tie, ${\bf v}'_i$ is chosen at random.

Evidently, $\eta$ represents the noise intensity; for $\eta= 2/3$ all information about the majority velocity is lost, rendering alignment impossible.  In the fully occupied lattice, naturally, no movement is possible, and the system becomes a 3-state majority-vote model \cite{zubillaga2022}, admitting an ordered phase for $\eta < \eta_c$ as discussed below. The fully occupied lattice allows neither 
particle displacements nor density fluctuations, and so cannot be considered ``active matter" \cite{chate2020b}. We nevertheless study the order-disorder transition at full occupancy since it is a limiting case of the model.
In the limit of vanishingly small density, on the other hand, each particle executes a persistent random walk, with a persistence time of $1/ \eta$ for small $\eta$; the walk is fully random for $\eta = 2/3$.

The model defined above bears certain similarities to the {\it four-state} square-lattice active Potts model proposed by Peruani {\it et al.}~\cite{peruani2011}. Like the model studied here, it also has excluded-volume interactions.  Aside from the
different lattice structure and number of allowed velocities, the model studied in \cite{peruani2011} features a different approach to the velocity-update process compared with our model or the original Vicsek model.  Specifically, the probability that a particle
change its velocity from ${\bf v}$ to ${\bf v}'$ is proportional to a Boltzmannlike factor:

\begin{equation}
    \label{eq:TR}
    \mbox{Prob} \left[{\bf v} \rightarrow {\bf v}'\right] \propto \exp \left( g \sum_{j \in A} {\bf v}'\! \cdot \! {\bf v}_j \right),
\end{equation}

\noindent where the sum is over the occupied nearest-neighbor sites of the particle under consideration and $g \geq 0$ is a parameter that plays the role of an inverse temperature. (Thus, although there is no one-to-one mapping, $\eta = 0$ in our model corresponds to $g \to \infty$ in the model of Peruani {\it et al.}, while $\eta = 2/3$ effectively corresponds to $g=0$.)  It is unclear, {\it a priori}, what effect such differences between the models might imply for the phase diagram.  
A further {\it procedural} difference is that Peruani {\it et al.} use random initial configurations (ICs) whereas, as explained in Sec. IIIA, we investigate several kinds of IC (including random positions and velocities) before concentrating our study on a particularly stable class of IC that we call an
{\it immobile band}. In light of the above considerations, the set of phases found in the two models need not be identical.

\section{\label{sec:result} Results and discussion}

\subsection{\label{subsec:resConf} Classification of steady-state configurations: preliminary survey}

\begin{figure}[!htp]
    \center
    \subfigure[a][]
    {\includegraphics[width=0.40\linewidth]{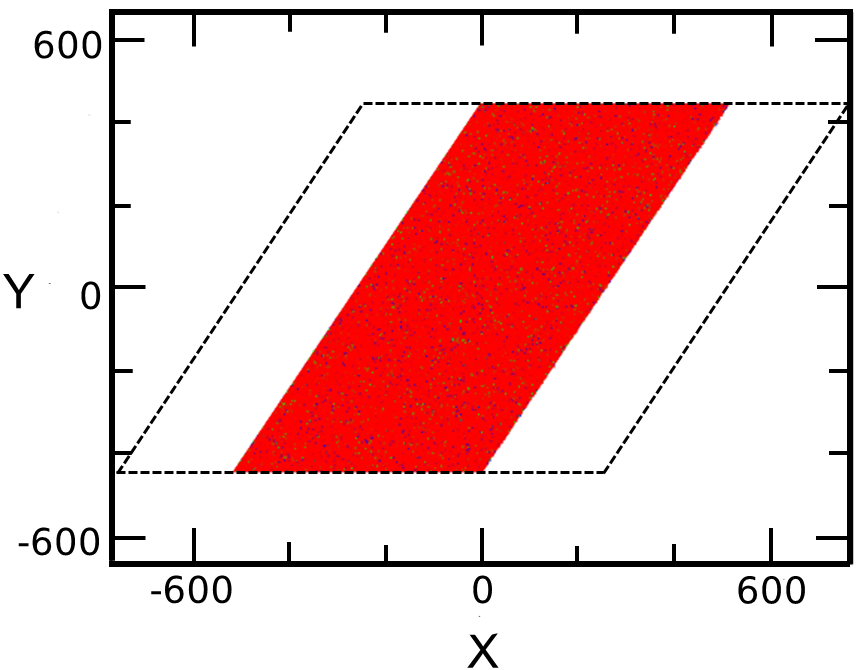}}
    \qquad
    \subfigure[b][]
    {\includegraphics[width=0.40\linewidth]{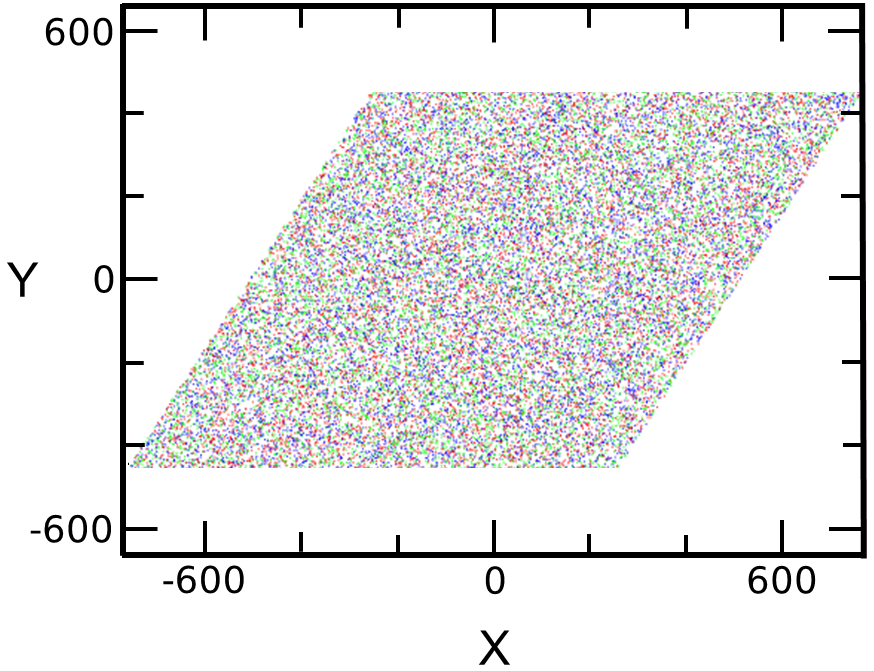}}
    \subfigure[c][]
    {\includegraphics[width=0.40\linewidth]{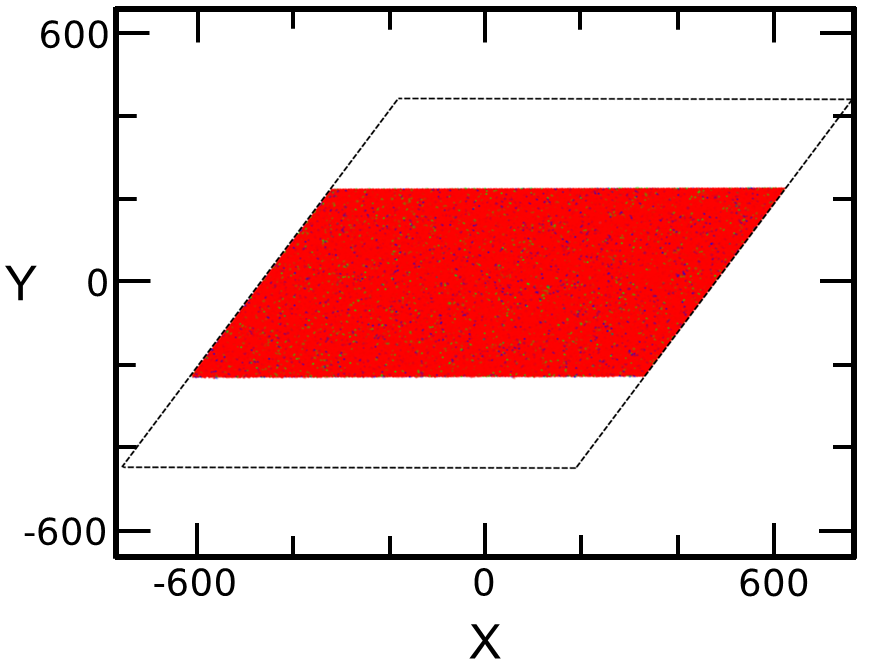}}
    \caption{\label{fig:confinit} (Color online) Examples of the three kinds of initial configuration (IC) used in this study. (a) Transverse band (TB), (b)  Random (R), (c) Immobile band (IB). While these images are for $L = 1025$ and $\rho = 0.5$, ICs are qualitatively similar for other sizes and densities. In random ICs, particles are assigned velocities independently, with equal probabilities from the set \{$ \hat{v}_1, \hat{v}_2, \hat{v}_3 $\}, and random positions, respecting the excluded-volume condition.  In both immobile- and transverse-band ICs, the particles are close-packed in a band that wraps the system, and whose width depends on the desired density.  In both IB and TB ICs, all particles have the same orientation.}
\end{figure}

The three-state active lattice gas exhibits diverse stationary configurations, including some previously observed in other models of active matter \cite{solon2013,peruani2011,karmakar2022,chate2008a,chate2008b}. In a set of preliminary studies extending to $10^5$ MC steps,
we simulated systems of size $L = \{33, 65, 129, 257, 513\}$, using three types of initial configuration (IC): Immobile Band (IB), Transverse Band (TB), and Random (R) (see Fig.~\ref{fig:confinit}). In these studies, the number of independent realizations varies from a minimum of five to a maximum of sixty for parameters that appear to place the system near a phase transition, that is, regions that exhibit a nonunique final configuration type and/or large fluctiuations in stationary properties such as the order parameter (see below). Study of the transverse band IC is motivated by previous works \cite{chate2008a, ginelli2016}, which indicate that in continuous-space models without volume exclusion (e.g., the Vicsek model) coexistence between ordered and disordered phases
is characterized by dense, ordered bands propagating perpendicular to the global velocity. Our results nevertheless indicate that, due to excluded volume, ordered configurations often consist of one or more {\it immobile} bands, motivating our study of ICs consisting of a single such band.

 The final configurations fall into six categories: Immobile Bands (IB); Type-1 and Type-2 Traffic Jams (TJI and TJII, respectively); Type-1 and Type-2 Mobile Bands (MBI and MBII, respectively); and Disordered Aggregates (DA). This classification, modeled on that employed by Peruani {\it et al.} \cite{peruani2011}, groups stationary states according to the shape and mobility of the flocks and/or particles. While the DA, IB, and TJI states observed here are similar to those reported in \cite{peruani2011}, we also find a second traffic-jam state (TJII), formed by the clash between two bands with distinct directions of motion. To the best of our knowledge, the MB states found here have been not observed previously.  On the other hand, the so-called {\it glider}
states reported in \cite{peruani2011} are not observed here.

\begin{figure}[!htp]
    \center
    \subfigure[a][] {\includegraphics[width=0.30\linewidth]{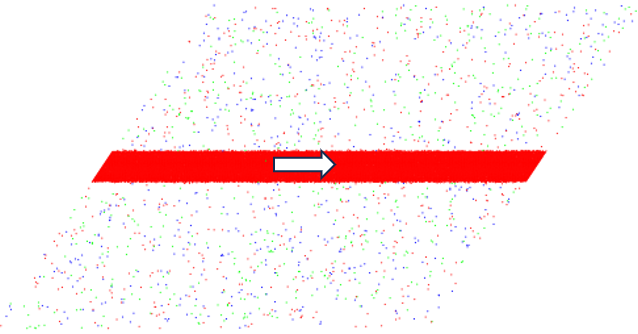}}
    \subfigure[b][] {\includegraphics[width=0.30\linewidth]{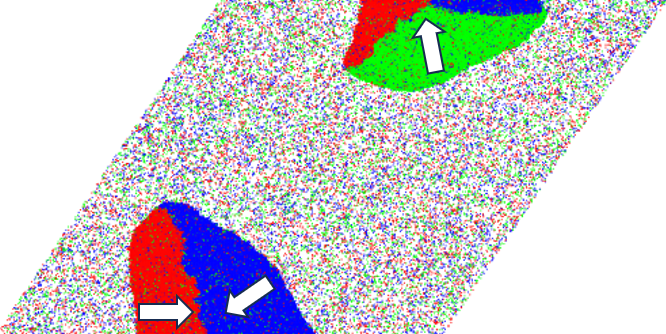}}
    \subfigure[c][]{\includegraphics[width=0.30\linewidth]{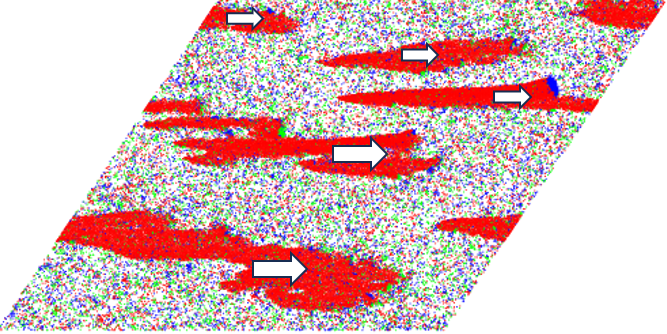}}
    \subfigure[d][] {\includegraphics[width=0.30\linewidth]{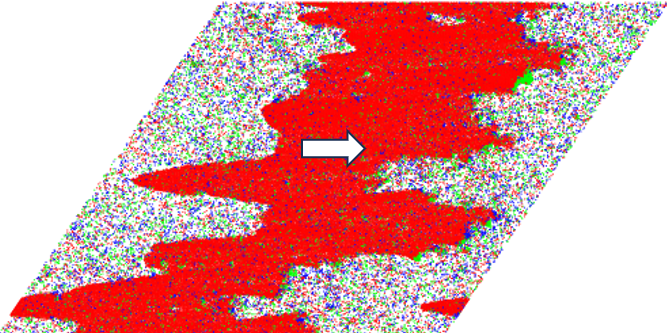}}
    \subfigure[e][] {\includegraphics[width=0.30\linewidth]{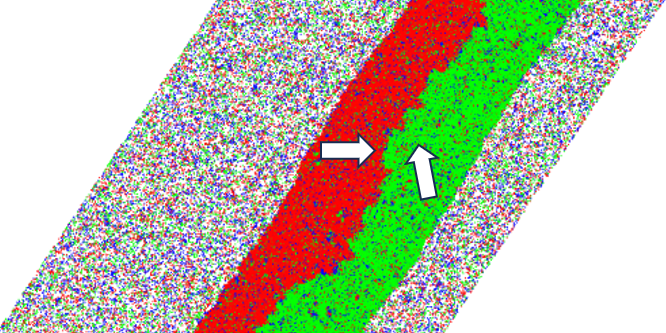}}
    \subfigure[f][] {\includegraphics[width=0.30\linewidth]{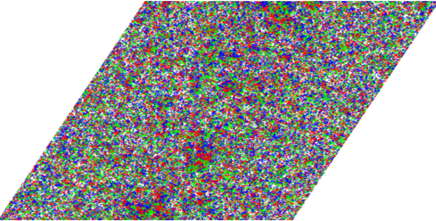}}
    \caption{\label{fig:typesteadystates} (Color online) Examples of final configurations observed after $10^5$ MC steps. (a) Immobile band (IB): $\rho=0.1$, $\eta=0.20$; (b) Type-I Traffic Jam (TJI): $\rho=0.3$, $\eta=0.42$; (c) Type-I Mobile Band (MBI): $\rho=0.3$, $\eta=0.38$; (d) Type-II Mobile Band (MBII): $\rho=0.5$, $\eta=0.4$; (e) Type-II Traffic Jam (TJII):  $\rho=0.5$, $\eta=0.44$. (f) Disordered Aggregate (DA):  $\rho=0.5$, $\eta=0.458$.  All images are for system size $L=513$. The white arrows show the majority velocity in each condensed region.}
\end{figure}

Figure~\ref{fig:typesteadystates} shows examples of each type of final configuration. Immobile-band states are similar in form to the eponymous ICs, except that in some cases (when using Random or TB ICs) two or more parallel bands are observed. Most of the particles in an IB are oriented parallel to the band, and so are blocked from moving forward by the particle immediately ahead. As a result, activity in this state is restricted mainly to band edges. At full occupancy, the ordered phase possesses a nonzero average
orientation, and may be seen as an IB occupying the entire lattice.

Non-IB condensed states typically occur at higher noise intensities than IBs. In the case of MBI, the velocities tend to align along the band, but small, short-lived flocks with different velocities appear, generating temporary congestion which is sufficient to prevent the formation of immobile bands, but not to eliminate global order entirely. Particles in these smaller groups tend to align, creating irregular clusters that move through space. A similar process occurs in MBII states, but here a band with irregular edges forms along a direction different from the direction of propagation, wrapping the system. 
Figure~\ref{fig:typesteadystates} suggests that in mobile bands, rough band edges allow particle movement to coexist with overall ordering along the band direction.

Type-II mobile bands are slightly reminiscent of the bands observed in Vicseklike models without excluded volume \cite{gregoire2003,chate2008b}. Nevertheless, the MBII
structures observed here have wildly fluctuating boundaries and a jammed bulk, compared to the smooth density variation and uniform motion of the bands observed without excluded volume.

Excluded volume also gives rise to ``traffic jams'' (TJI and TJII), congested configurations in which flocks of particles with different velocities block each other's motion. In TJI states, two or more clusters with different velocities meet head-on, leading, in general, to an oval structure. This configuration occurs through the growth of clusters with different velocities; typically MBIs are precursors. In this type of configuration, the alignment interaction leads to competition between groups with different velocities to ``capture'' new particles, and so to a dynamic process of evaporation/condensation.
By contrast, TJIIs emerge at higher densities ($\rho \geq 0.5$) and consist of two wrapping stripes with different velocities that again block each other's motion. Such configurations exhibit a slower evaporation/condensation process than TJI, suggesting that TJII configurations have a longer lifetime than do TJIs.

Finally, disordered steady states, characterized as ``disordered aggregates" (DA) are observed, as expected, at low densities and high noise intensities.  At densities $\rho \leq 0.96$, condensed structures, be they bands or traffic jams, {\it coexist} with a low-density, disordered ``vapor".  
In fact, the vapor and the homogeneous DA state are one and the same phase.  This conclusion is based on studies of the radial distribution and velocity correlation functions described below.

\begin{figure}[!htp]
    \center
    \subfigure[a][] {\includegraphics[width=0.46\linewidth]{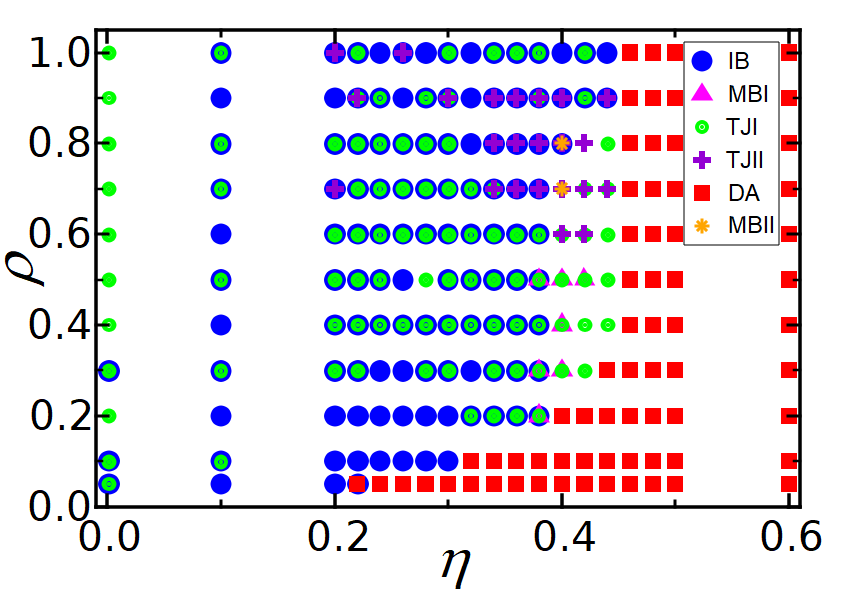}}
    \qquad
    \subfigure[b][] {\includegraphics[width=0.46\linewidth]{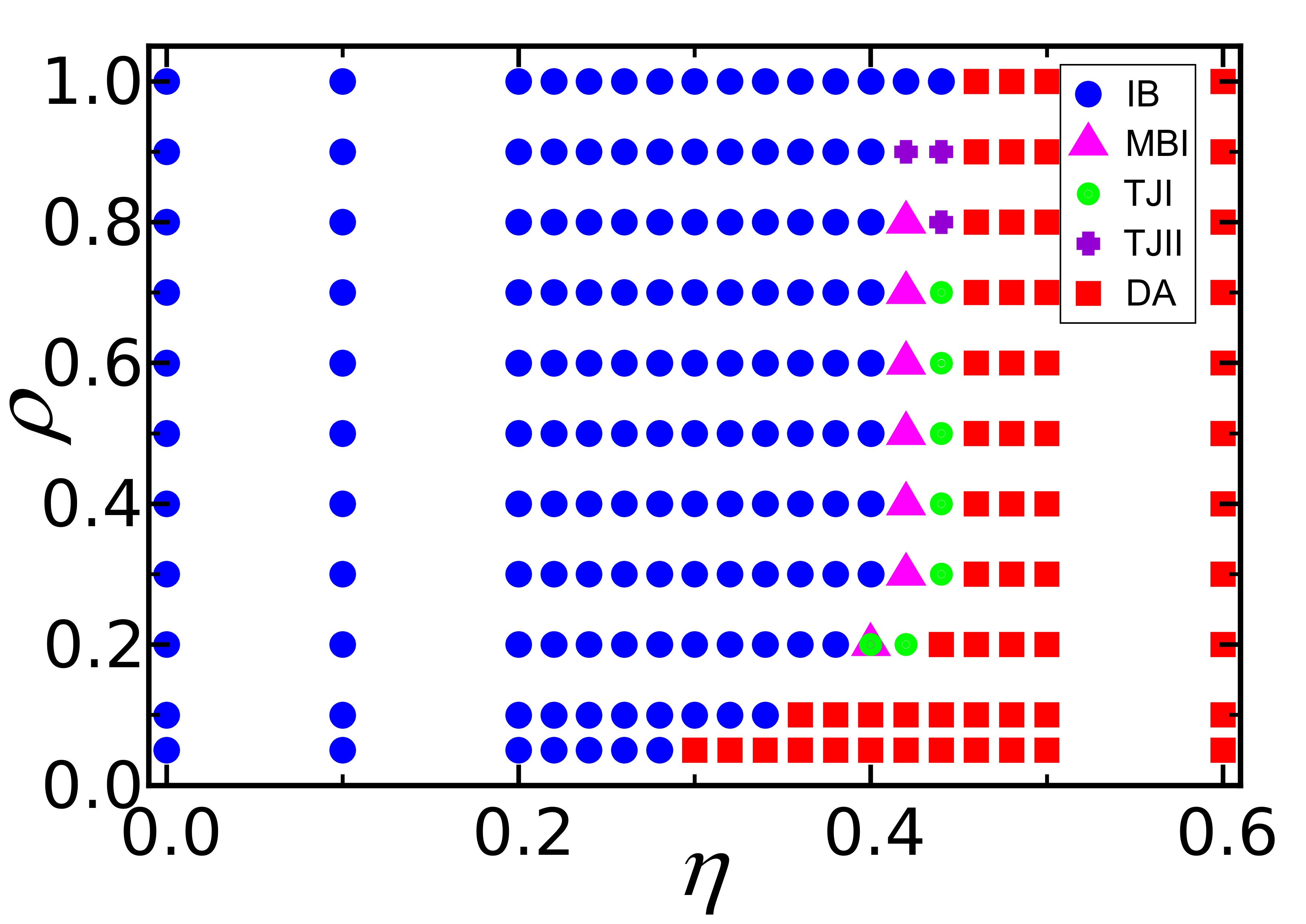}}
    \newline
    \subfigure[c][]{\includegraphics[width=0.46\linewidth]{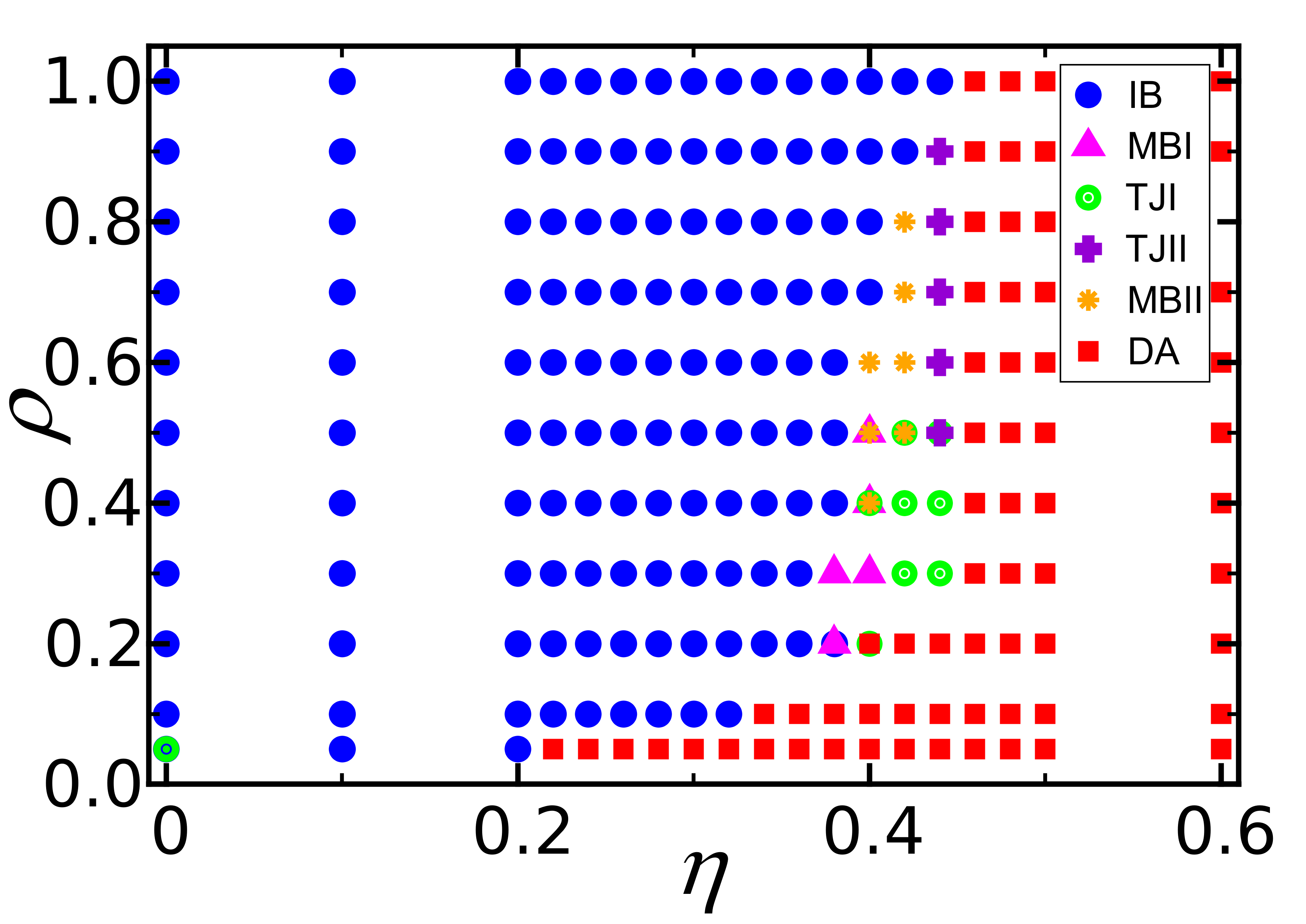}}
    \qquad
    \subfigure[d][]{\includegraphics[width=0.47\linewidth]{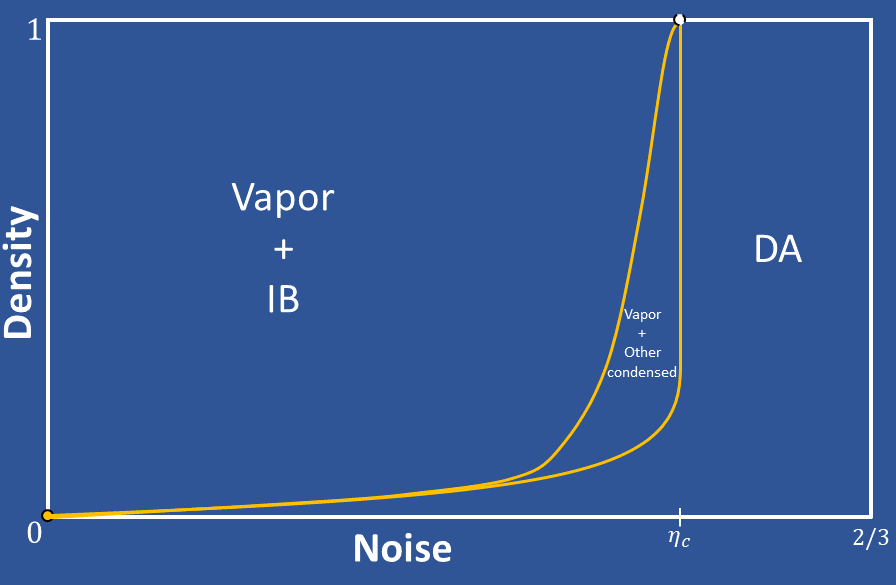}}
    \caption{\label{fig:mapConfigurationL513} (Color online) Maps of the occurrence of the states for $L = 513$ and initial configurations as noted. (a) Random, (b) IB, (c) TB. (d) Generic phase diagram at ($\rho$, $\eta$) plane. Note that in some cases, more than one kind of final state is possible at the same point in parameter space, particularly for random ICs.}
\end{figure}

Our first objective is to infer the phase diagram in the $\rho-\eta$ plane (i.e., in the infinite-$L$ limit) based on maps of the occurrence of the states shown in Fig.~\ref{fig:typesteadystates}. 
To begin, we determine which regions of parameter space typically yield each type of configuration. While the fundamental intensive parameters are density $\rho$ and noise intensity $\eta$, the boundaries between regions exhibiting one or another configuration type also vary systematically with the system size $L$. Figure~\ref{fig:mapConfigurationL513} shows where each configuration is observed using a relaxation time of $10^5$ MC steps.
(Similar maps for $L = 33$, $65$, $129$ and $257$ are provided in the Supplemental Material \cite{supp}.)

Examination of Fig. \ref{fig:mapConfigurationL513} reveals that at certain points in the $\rho - \eta$ plane, more than one kind of configuration can arise from a given IC; this multiplicity is most common for random ICs. Regardless of the initial configuration or the density, the final state is DA for sufficiently large noise intensity ($\eta \gtrsim 0.45$).  The value of $\eta$ marking the transition between DA and an ordered state increases rapidly with $\rho$ at low densities and saturates as $\rho \to 1$. 

As noted above, for random initial configurations, a greater variety of condensed states are observed, including nonunique outcomes for the same parameters.  This is particularly common for larger systems, suggesting that the evolution becomes trapped in a metastable state with a lifetime that grows with system size.  A similar observation holds for TB ICs.  The hypothesis that metastability is responsible (at least in part) for the variety and nonuniqueness of steady states is supported by the observation that, even when using a TB initial configuration, MBII steady states are only observed in larger systems, for which the lifetime presumably exceeds the simulation time of $10^5$ MC steps. We return to this issue in the following subsection.

While Random or TB initial configurations can, as noted, lead to multi-IB states, the number of stripes observed varies with each realization and tends to shrink with increasing noise intensity $\eta$. For example, for random ICs, $L = 513$, $\rho  = 0.1$ and $\eta = 0.1$, we observe as many as fifteen bands, whereas near the phase transition ($\eta \approx 0.32$) only a single band is found.

\subsection{\label{subsec:resConfdetail} Steady states: detailed study}

The results reported in the previous subsection represent a preliminary step in characterizing the phase diagram.  To obtain a more precise picture, we use longer simulation times ($10^7$ MC steps) and restrict attention to IB initial configurations, due to the relative simplicity and apparent stability of IB steady states.  We stress that for IB initial configurations, which employ a single band, multiple-IB steady states are never observed.  Limited computational resources prohibit our repeating all of the $\eta$ values and system sizes investigated for $10^5$ MC steps in these more detailed studies.

\begin{figure}[!htp]
    \centering
    \includegraphics[width=0.75\linewidth]{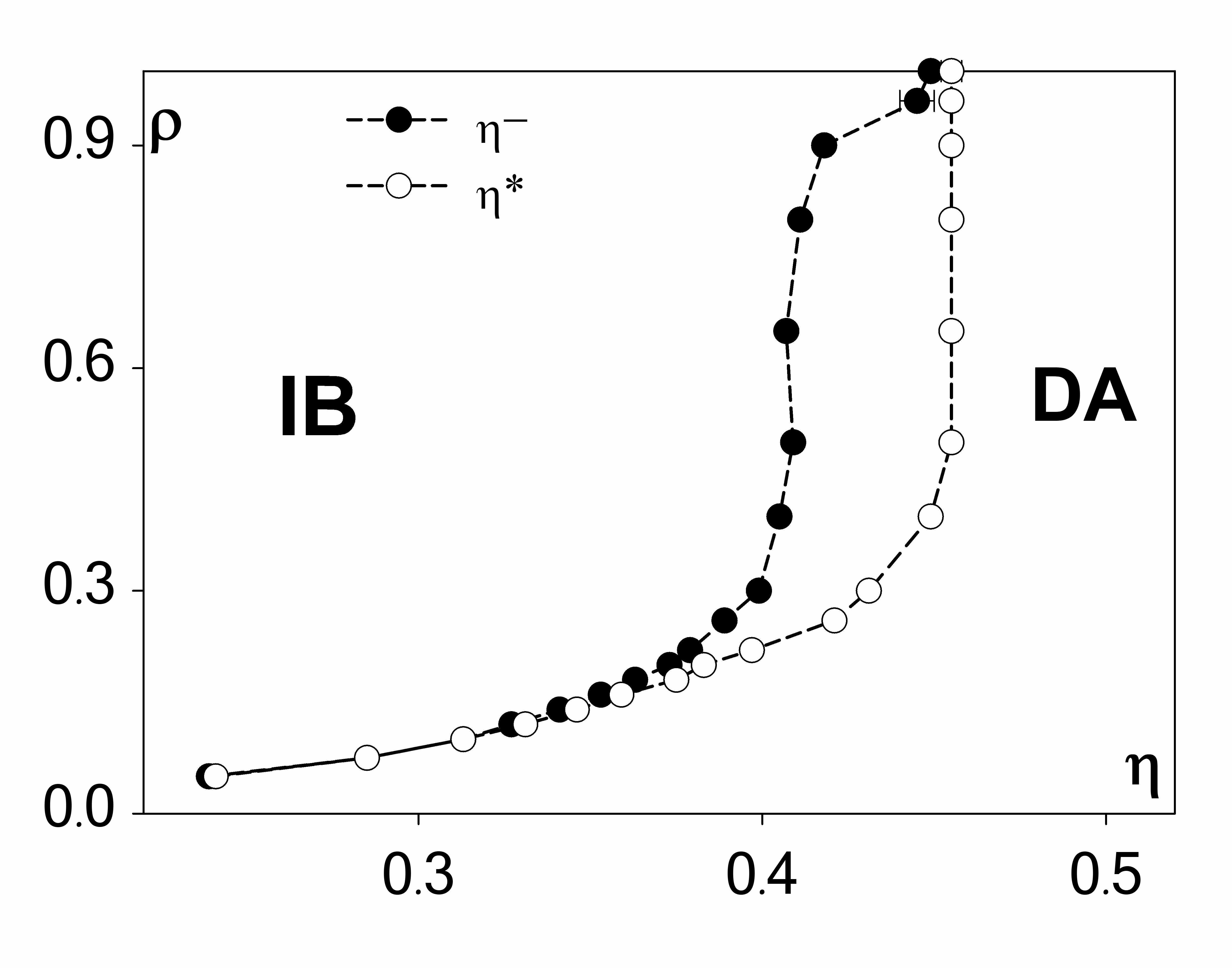}
    \caption{\label{fig:eta*eta-} Stability limits $\eta^-$ and $\eta^*$ versus $\rho$ for system size $L = 129$, and a simulation time of $10^7$ MC step. Dashed lines connecting the points are merely an aid to visualization.}
\end{figure}

We begin by probing the stability limits of the IB and DA phases.  Let $\eta^- (\rho,L)$ 
denote the noise intensity below which the stationary configuration is {\it always} IB.
Similarly, define $\eta^*(\rho,L)$ as the noise intensity above which only the DA phase is observed. Figure~\ref{fig:eta*eta-} shows these stability limits for $L = 129$ and a simulation time of $10^7$ MC steps, and is in qualitative agreement with Fig.~\ref{fig:mapConfigurationL513}(d). 
In the region between the two stability limits, other condensed states can appear, depending on the density. We note however that MBII configurations are not observed in studies extending to $10^7$ MC steps, and so are excluded from the set of phases.  All the other configuration types found for the survey using $10^5$ MC steps are also observed in the longer studies; we shall regard them as phases of the model.  A systematic determination of the MBII lifetime, or of inter-phase switching times (see below) is beyond the scope of the present work.

The extended simulations provide clear evidence of bistability 
in the parameter space between the stability limits, vide Fig.~\ref{fig:eta*eta-}, in that the configuration may alternate between phases over time. For lower densities, we
find time series in which IB and DA phases alternate. At higher densities, we find 
time series with alternation between the following pairs of phases: IB-DA, IB-MBI, IB-TJI, MBI-TJII, TJI-DA, MBI-DA, TJI-TJII, and, most commonly, MBI-TJI. 

\begin{figure}[!htp]
    \center
    \subfigure[a][] {\includegraphics[width=0.46\linewidth]{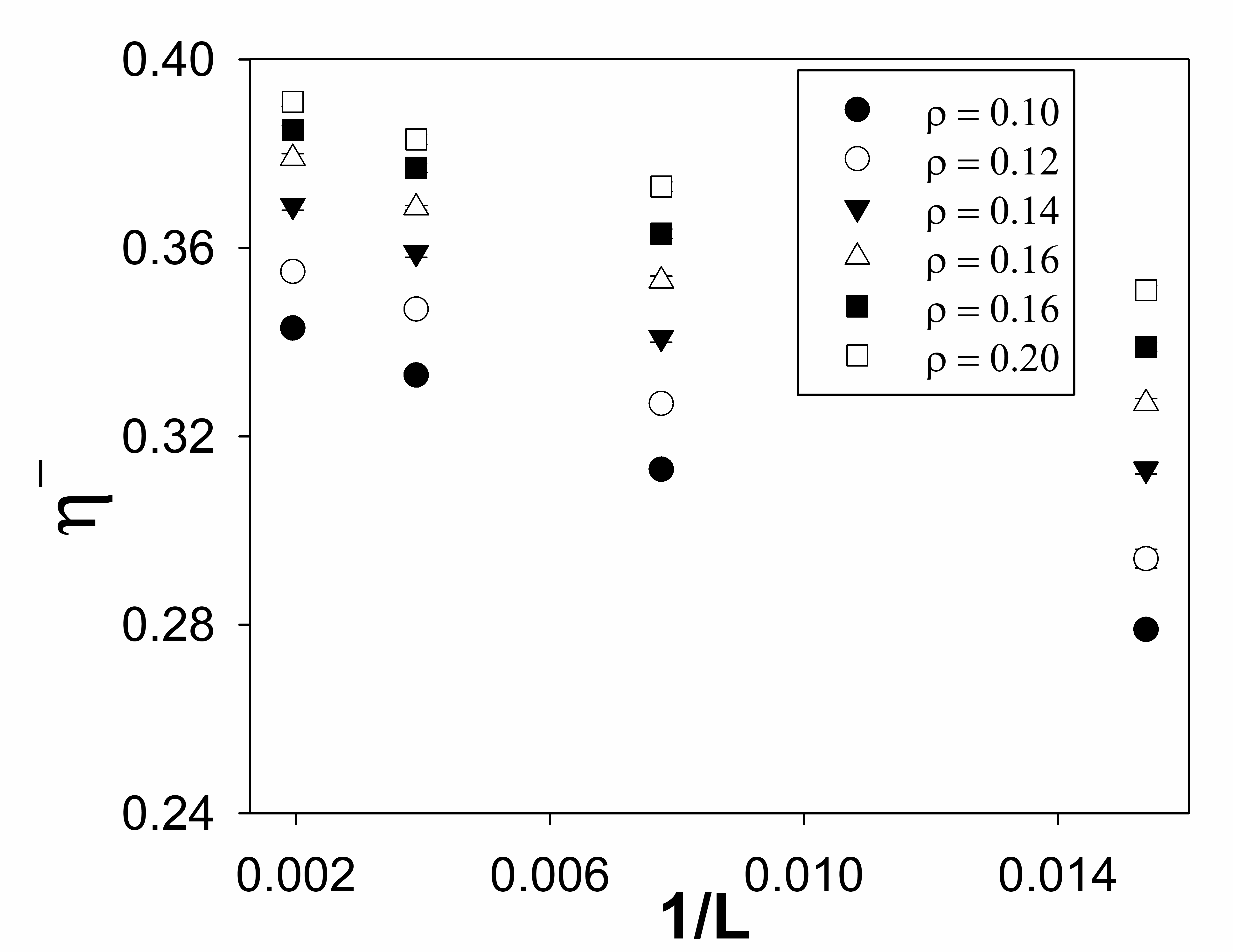}}
    \qquad
    \subfigure[b][] {\includegraphics[width=0.46\linewidth]{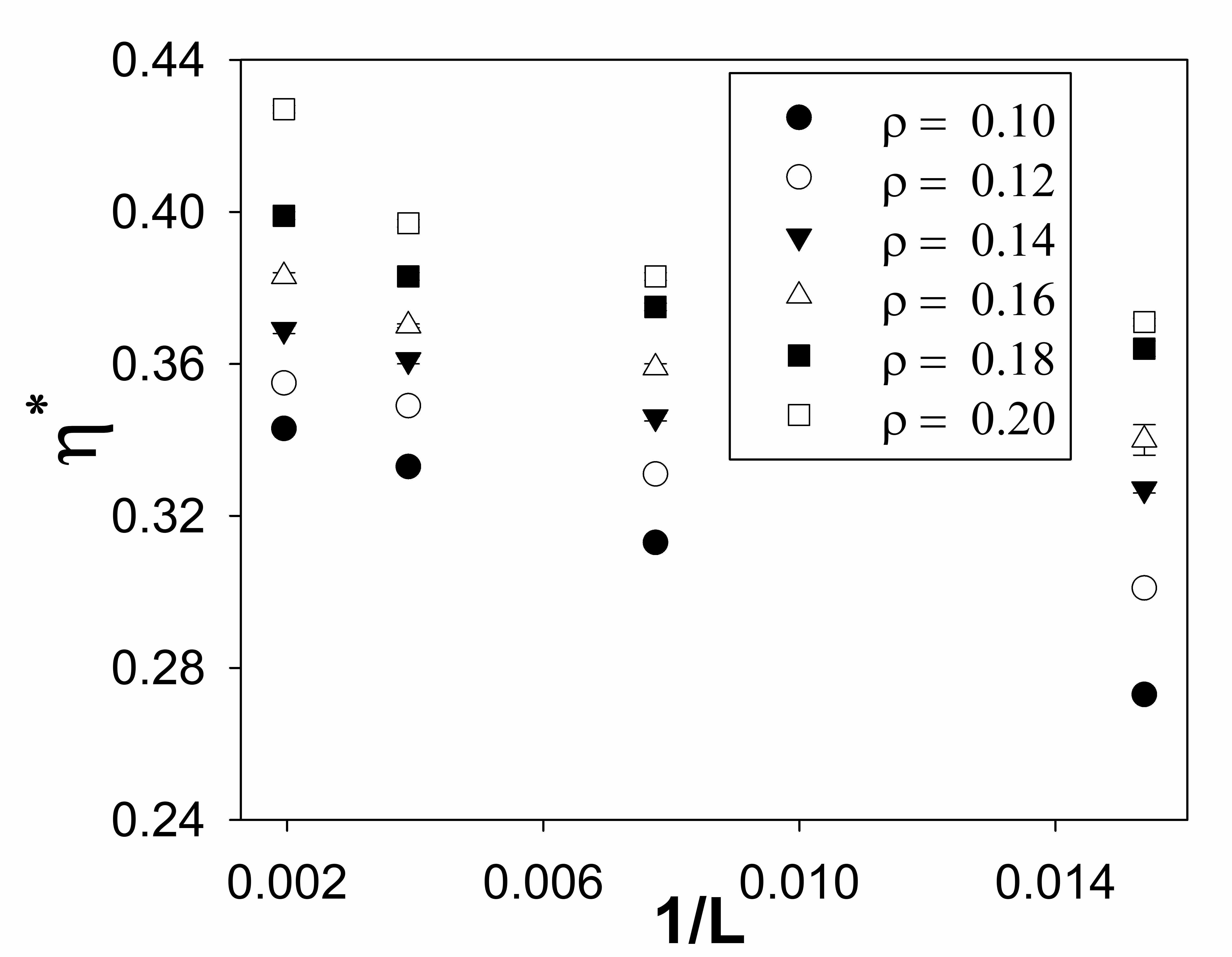}}
    \newline
    \subfigure[c][]{\includegraphics[width=0.46\linewidth]{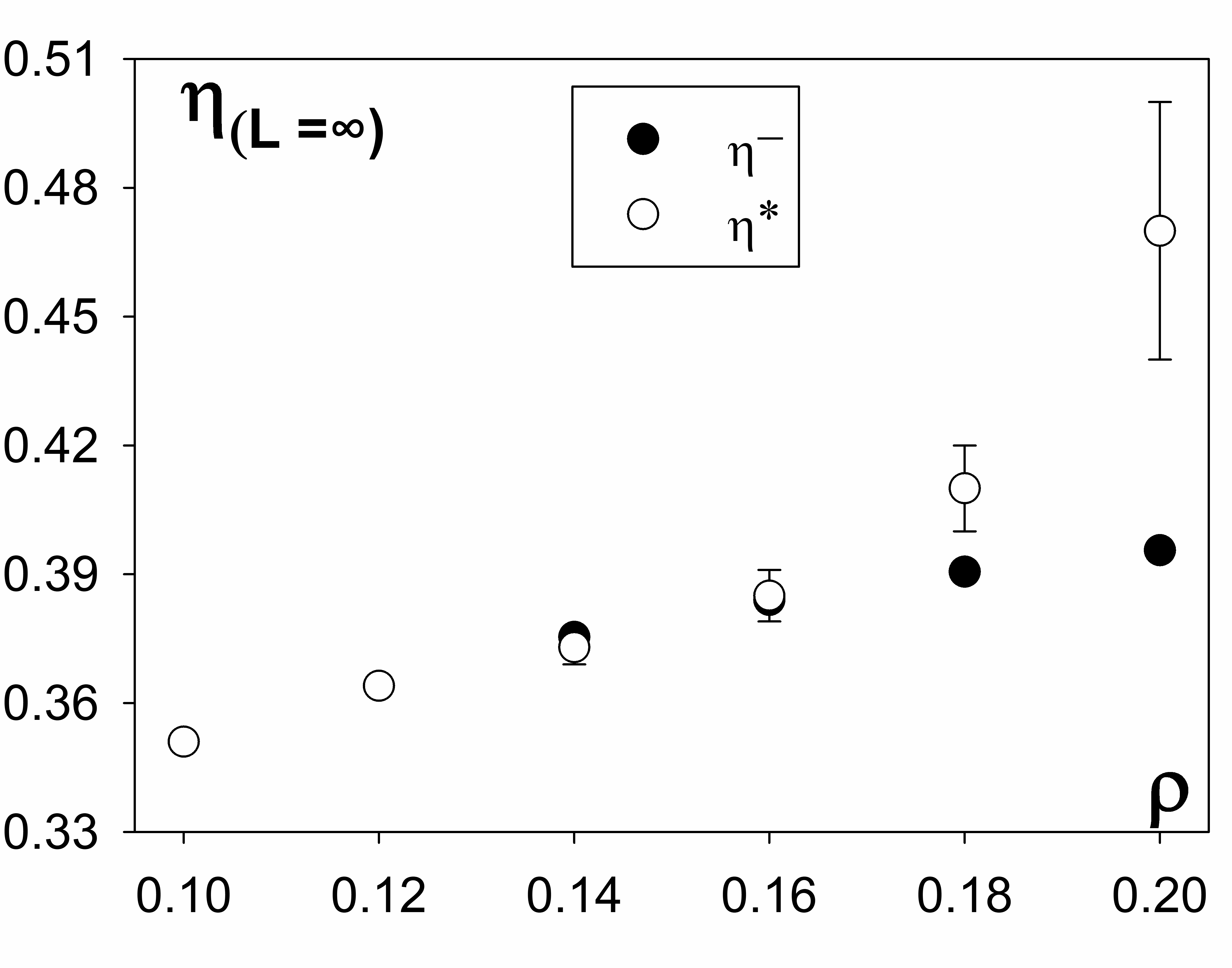}}
    \caption{\label{fig:limtTermo} (a) $\eta^-$ versus $1/L$ at low densities; (b) $\eta^*$ versus $1/L$; (c) Estimated values of $\eta^-$ and $\eta^*$ in the thermodynamic limit.}
\end{figure}

Figure~\ref{fig:limtTermo}(a)-(b) shows $\eta^-$ and $\eta^*$ as a function of system size in the low-density regime. The values for $\eta^-$ are well fit by a straight line. On the other hand, $\eta^*$ exhibits positive curvature with increasing density. We believe that this curvature is associated with the fact that, for these higher densities, the system goes through several states during the transition from IB to DA, as we increase the noise intensity. The intermediate states (MBI, TJI, and TJII)  
are readily nucleated at higher densities, thus
extending the stability of condensed phases to higher noise intensities.
Figure~\ref{fig:limtTermo}(c) shows the extrapolated values for $\eta^-$ and $\eta^*$, which again indicates that the phase diagram is 
qualitatively as suggested in Fig.~\ref{fig:mapConfigurationL513}(d).  The gap
between the IB and DA phases is seen to extend to $\rho = 0.95$ (the highest density studied), suggesting that the order-disorder transition is discontinuous for {\it any}
density smaller than unity.

\begin{figure}[!htp]
    \centering
    \includegraphics[width=0.85\linewidth]{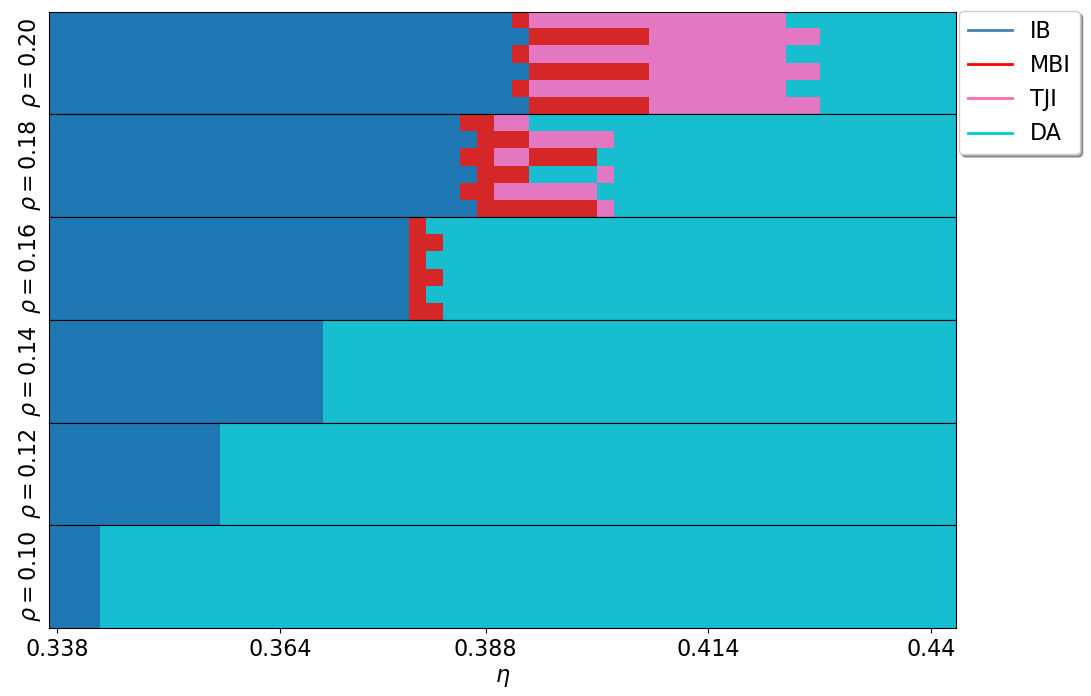}
    \caption{\label{fig:digramL513} (Color online) Observed states for noise levels near the order-disorder transition at low density and $L = 513$. Note that the number of times a configuration is indicated at a certain point in configuration space is not related to its frequency of occurrence.}
\end{figure}

To gain a better understanding of the transition region at low densities ($0.1 \leq \rho \leq 0.2$), we perform simulations for $L = 65$, 129, 257, and 513, again using $10^7$ MC steps, and a density increment of $\Delta\rho=0.02$. Figure~\ref{fig:digramL513} illustrates the observed states for $L = 513$. It is evident that as density increases, the number of possible states also grows.  At several points exhibiting nonunique final configurations, the system alternates between states, confirming bistability, as described above. Far from transition regions, the states are unique. In general, the sequence of states observed with increasing noise is IB, MBI, TJI, and DA.

Type-2 traffic jams are found at higher densities, presumably because they involve two bands that wrap the system and so require more particles to remain stable.
Studies using $L = 129$, densities $\rho \geq 0.5$, and noise intensities close to $\eta^*$, yield TJII final states. These transitions are often governed by an evaporation/condensation process alternating with the TJI state but persisting over the duration of the simulation. Analyses for other system sizes yield qualitatively similar behavior.

We close this subsection with a remark on rather exotic ordered, spatially uniform
states with density smaller than unity.
Consider, for example, a configuration in which all sites in one of the four sublattices of the triangle lattice are occupied by particles having the same velocity (say, $v_1$) and all other sites are vacant.  For $\eta = 0$, this IC evolves to a state in which, on average, half the particles in each occupied row are mobile.
[Note that for zero noise the system is in fact a stack of independent rings, each
running a totally asymmetric exclusion process (TASEP).]
Studies of a system with $L=129$ extending to $10^8$ MC steps show that this state is unstable to arbitrarily small nonzero noise intensities.  As one increases $\eta$, the final state passes through the sequence: multiple IBs, single IB, TJI and DA.

\subsection{\label{coexistence} Coexistence between condensed phases and DA}

As noted above, for densities smaller than unity, condensed phases
coexist with a low-density vapor.  Here we show that the properties of the vapor
are those of the disordered aggregate (DA) phase, as characterized by the
radial distribution function $g(r)$ and velocity correlations.

\begin{figure}[!htp]
    \centering
    \subfigure[a][] {\includegraphics[width=0.655\linewidth]{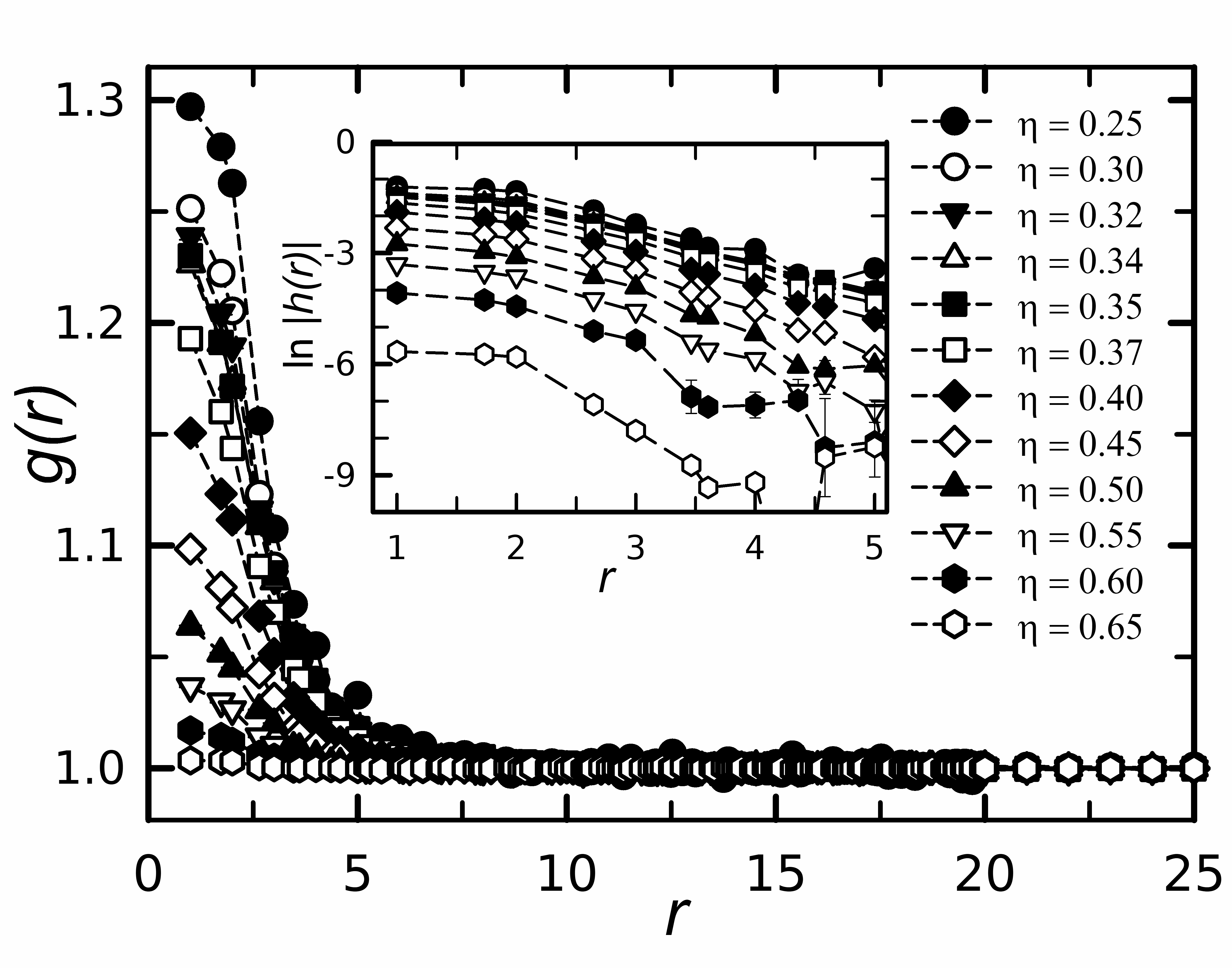}}
    \subfigure[b][] {\includegraphics[width=0.49\linewidth]{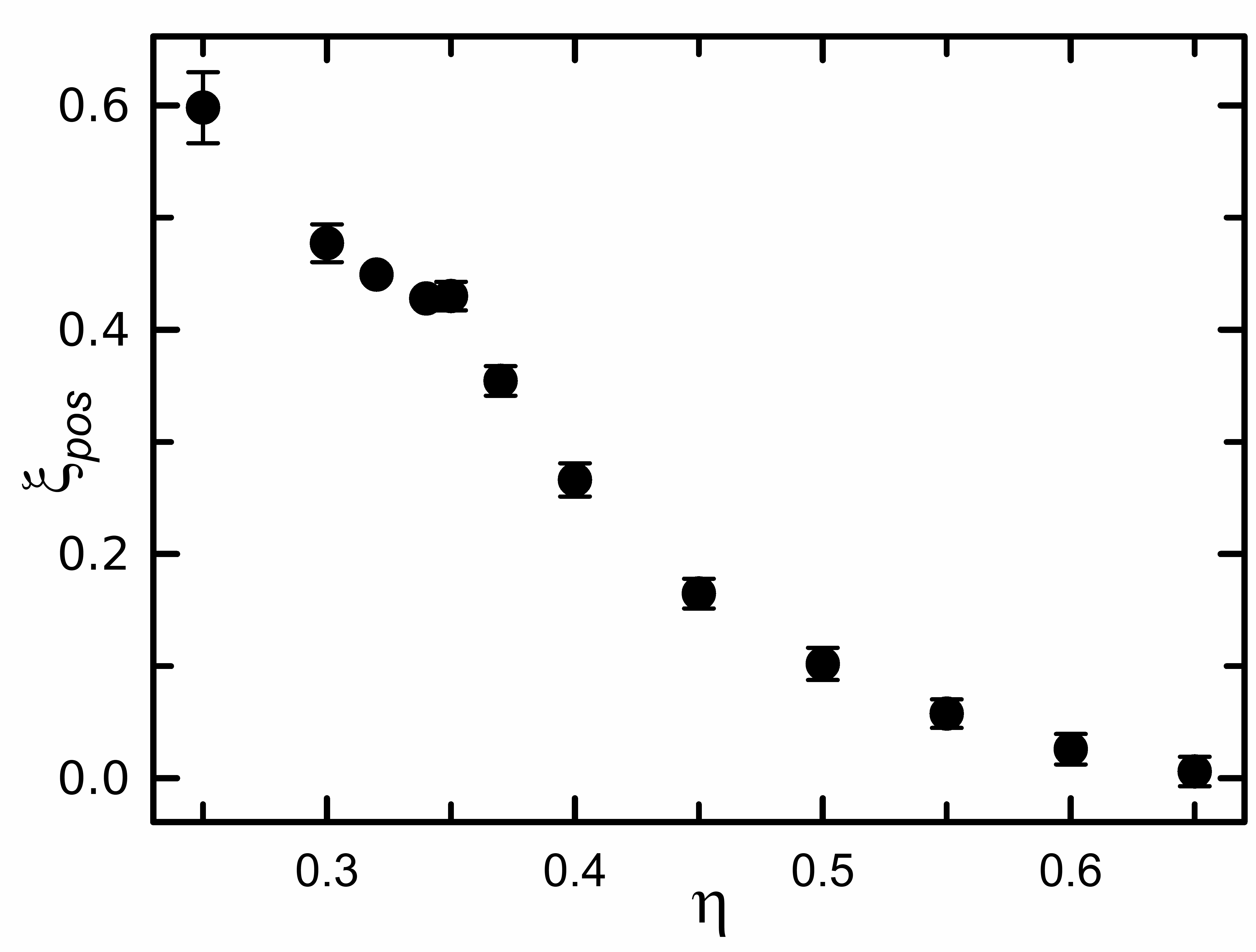}}
    \subfigure[c][] {\includegraphics[width=0.49\linewidth]{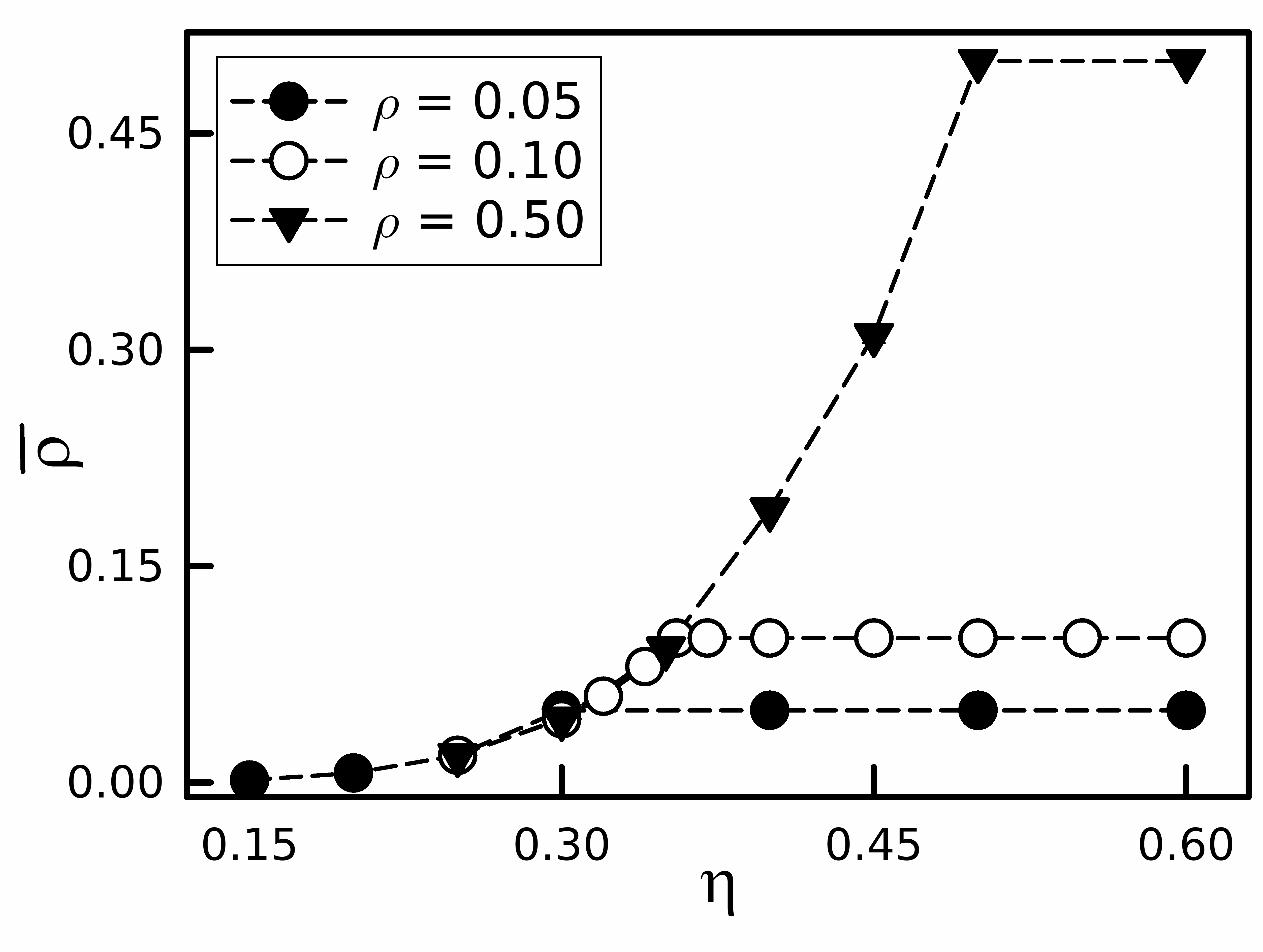}}
    \caption{ \label{fig:grL513} (a) $g(r)$ for $\eta$ values as indicated, in studies with $\rho = 0.1$ and $L=513$, including the vicinity of the IB/DA phase transition, which occurs at $\eta = 0.345(5)$ for this density. Inset: $\ln|h(r)|$ versus $r$; (b) Correlation length ($\xi_{pos}$) versus $\eta$ for the same parameter values as in (a). (c) $\overline{\rho}$ {\it vs} $\eta$ for $\rho = 0.05, 0.1$, and $0.5$.} 
\end{figure}

A convenient definition of $g(r)$ for lattice models \cite{scalas1994} can be written so:

\begin{equation}
\label{eq:defgr}
    g(r) = \frac{1}{\overline{N}} \sum^{\overline{N}}_{l=1}\frac{1}{\mathcal{N}_l(r)} \sum^{\overline{N}}_{m=1} \delta(r-|{\bf r}_m-{\bf r}_l|),
\end{equation}

\noindent where $\mathcal{N}_l(r)$ is the number of sites at a distance $r$ from the $l$-th particle, ${\bf r}_m$ and ${\bf r}_l$ are the positions of particles $m$ and $l$, respectively, and the sums are over all $\overline{N}$ particles in the vapor/DA phase. 
$\overline{\rho}$ corresponds to the density of the vapor and/or DA. (In the presence of a condensed phase, the vapor density is smaller than the overall density.) 
In brief, $\overline{\rho} g(r)$ represents the probability that site 
${\bf r}$ is occupied, given that there is a particle at the origin. 
Figure~\ref{fig:grL513}(a), for $L=513$ and $\rho=0.1$, shows that $g(r)$ decreases rapidly with $r$, 
eventually leveling off near unity for $r > 10$. 
Note that the function $g(r)$ varies smoothly with $\eta$, suggesting that the vapor that coexists with IB, and the DA, are the same phase.

The pair correlation function, $h(r) \equiv g(r)-1$, measures the extent to which the presence of a particle at the origin is correlated to the presence of another at a distance $r$; the inset of Fig.~\ref{fig:grL513}(a) shows that $h(r) $ decays in a roughly exponential manner.
Introducing the correlation length, $\xi_{pos} = \frac{1}{2} \sum_{n=1}^{n_{max}} |h(r_n)+h(r_{n+1})| (r_{n+1} - r_n)$ (where $\{r_1 = 1, r_2 = \sqrt{3}, r_3 = 2, ...\}$ are the first-,  second-, third-,... neighbor distances), Fig.~\ref{fig:grL513}(b) shows that $\xi_{pos}$ decreases continuously, monotonically (and quite rapidly) with increasing noise. (We note that at the low density considered in Fig.~\ref{fig:grL513}, the transition is discontinuous (see Sec. IIID), so that $\xi_{pos}$ remains finite.)
These findings again support the conclusion that vapor and DA are a single phase. 

The $\overline{\rho}$ increases with $\eta$ until saturating at $\overline{\rho} = \rho$, as seen in Fig.~\ref{fig:grL513}(c). The apparent discontinuity in $d \xi_{pos}/d \eta$ occurs at the IB/DA phase boundary, and can be attributed to the singular dependence of $\overline{\rho}$ on $\eta$ at this point.

We define the velocity (or orientation) correlation function, $C_v(r)$, as, 

\begin{equation}
    C_v (r) = \frac{\sum \limits_{l,m=0}^{\overline{N},\overline{N}} {\bf v}_l \cdot {\bf v}_m\;\delta(r-|{\bf r}_m-{\bf r}_l|)} {\sum \limits_{l,m=0}^{\overline{N},\overline{N}} \delta(r-|{\bf r}_m-{\bf r}_l|)},
    \label{eq:defCv}
\end{equation}

\noindent where  ${\bf v}_l$ , ${\bf v}_m$, ${\bf r}_l$ and ${\bf r}_m$ are velocities and positions of particles $l$ and $m$, respectively, and the sums are over all $\overline{N}$ particles in the vapor or DA phase.

\begin{figure}[!htp]
    \centering
    \includegraphics[width=0.80\linewidth]{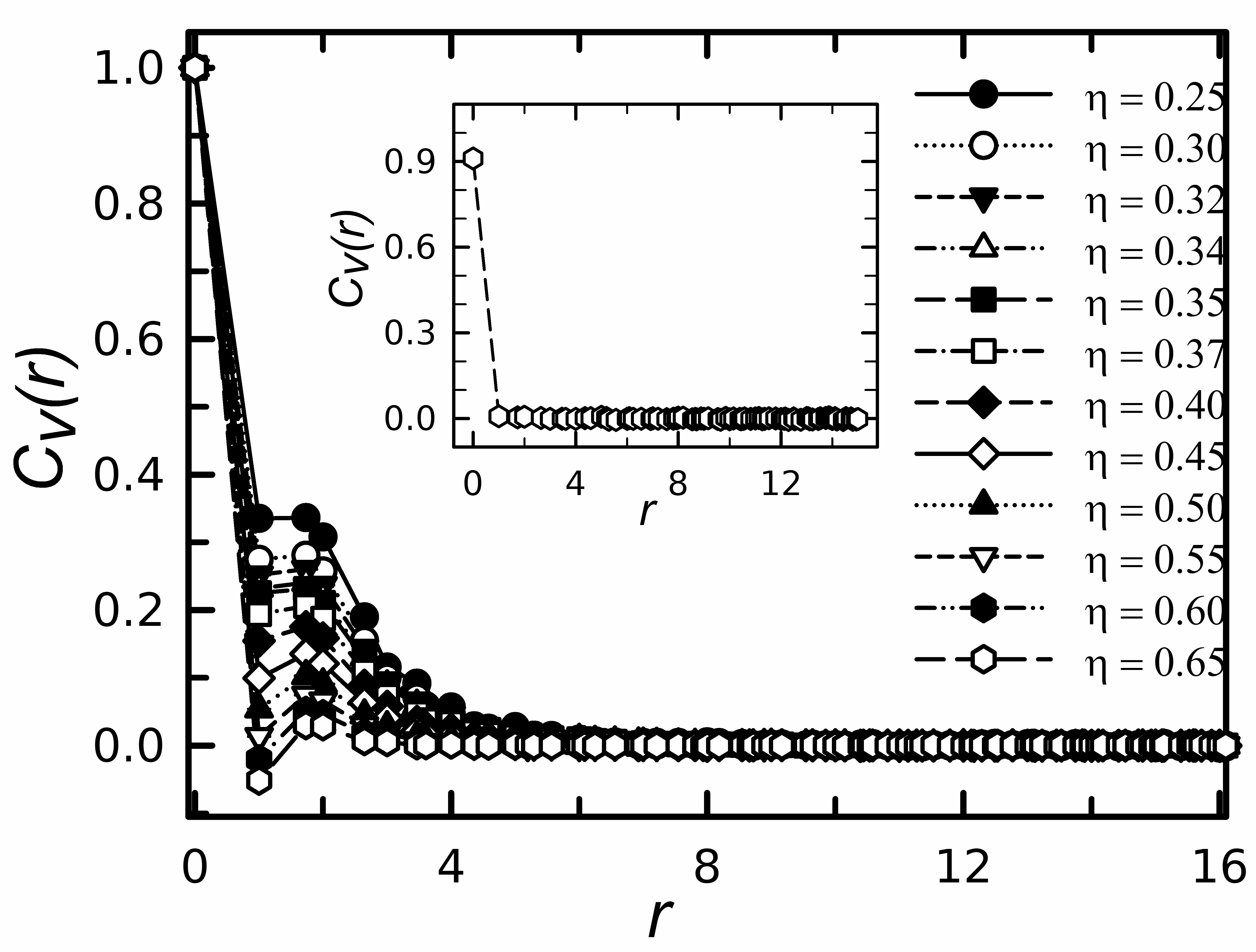}
    \caption{\label{fig:cvD01} $C_v(r)$ for noise intensities as indicated, $L = 513$, and $\rho = 0.10$. Inset: $C_v(r)$ for the model without excluded-volume interactions, using the same parameters, with $\eta = 0.65$.}
\end{figure}

Figure~\ref{fig:cvD01} shows that much like $g(r)$, the functions $C_v (r)$ change smoothly with $\eta$ in the vicinity of the IB/DA transition, again indicating that vapor and DA are the same phase. Curiously, in the limit of random reorientation ($\eta \simeq 2/3$), the velocities of particles occupying neighboring sites are {\it anticorrelated}, due to excluded volume. For these high noise levels and low densities, there are no condensed structures, but the excluded-volume interaction prohibits, for example, a particle at the origin with velocity ${\bf v}_1$ acquiring a neighbor with the same velocity at the neighboring site (1,0). Indeed, the inset of Fig.~\ref{fig:cvD01} shows that the anticorrelation vanishes in the absence of the excluded-volume interaction.

Analyses of $g(r)$ and $C_v(r)$ for $\rho = 0.05$, $0.2$, and $0.5$, yield comparable findings. Finally, we compare the properties studied above in vapor and DA phases having the same density at the same noise intensity, $\overline{\rho} = 0.0597$, and $\eta = 0.32$ (both with $L = 513$). The values of $g(r)$ and $C_v(r)$ are the same to within uncertainty in the two cases. Since significant differences in the properties of vapor and DA states are absent, we conclude that they constitute a single phase whose properties vary smoothly with $\eta$ and $\rho$.

\subsection{\label{subsec:resQuan} Order parameter}

As is customary in the study of active matter \cite{vicsek1995,ginelli2016},
we define the order parameter as the modulus of the mean velocity,

\begin{equation}
	\label{eq:order_parameter}
	\phi =  \frac{1}{N} \left| \sum_{i=1}^{N}{\bf v}_i\right|;
\end{equation}

\noindent where $N$ is the particle number and ${\bf v}_i$ is the velocity of particle $i$.

In the disordered phase, $\phi \rightarrow 0$ as the system size $L$ approaches infinity. We note that since $\phi$ is a measure of {\it global} order, it is not reliable in distinguishing, for example, between TJI and DA phases. The former features ordering of particles into high-density regions of comparable size, but with different directions of movement, so that $\phi$ may be close to zero in a typical configuration.

\begin{figure}[!htp]
    \center
    \subfigure[a][] {\includegraphics[width=0.49\linewidth]{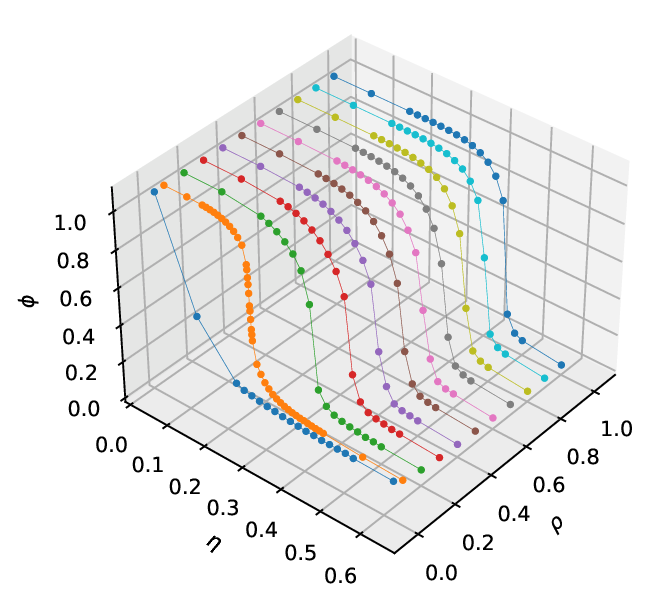}}
    \subfigure[b][] {\includegraphics[width=0.49\linewidth]{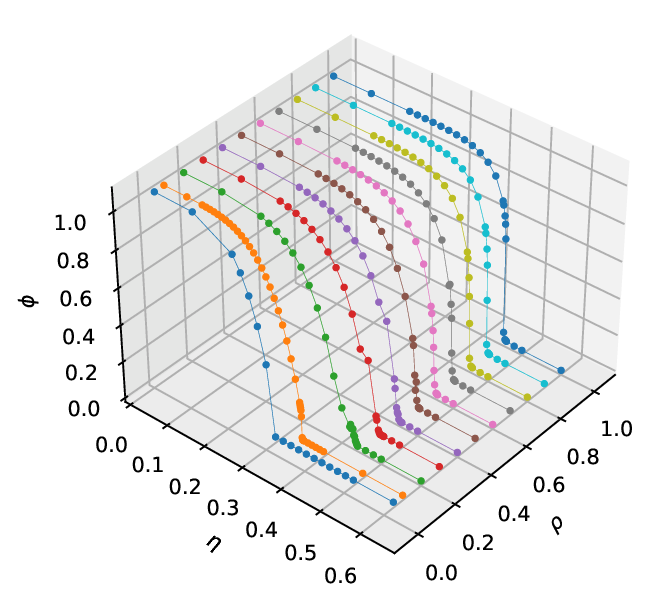}}
    \caption{\label{fig:phi3d} (Color online) Order parameter as a function of $\rho$ and $\eta$ for $L=33$ (left) and $513$ (right), and IB initial configurations. The densities are: 0.05, 0.1, 0.2, 0.3, 0.4, 0.5, 0.6, 0.7, 0.8, 0.9 and 1.0}
\end{figure}

Figure~\ref{fig:phi3d} shows $\phi$ as a function of $\rho$ and $\eta$ for $L=33$ and $513$, using IB initial configurations. Here, in each independent realization, the system is allowed to relax for $2 \times 10^4$ time steps, with $\phi$ obtained from a temporal average evaluated over the subsequent $8 \times 10^4$ time steps; we then evaluate the mean of $\phi$ over the set of independent realizations. As expected, in the zero-noise limit, the order parameter assumes its maximum value; for $\eta = 2/3$, particle velocities are chosen randomly, with equal probabilities, at each update, so that $\phi \rightarrow 0$ as $L \to \infty$. Regardless of the density $\rho$, as the noise intensity increases, there is an abrupt decrease in $\phi$ associated with the transition to the DA state.

\begin{figure}[!htp]
    \center
    \subfigure[a][] {\includegraphics[width=0.47\linewidth]{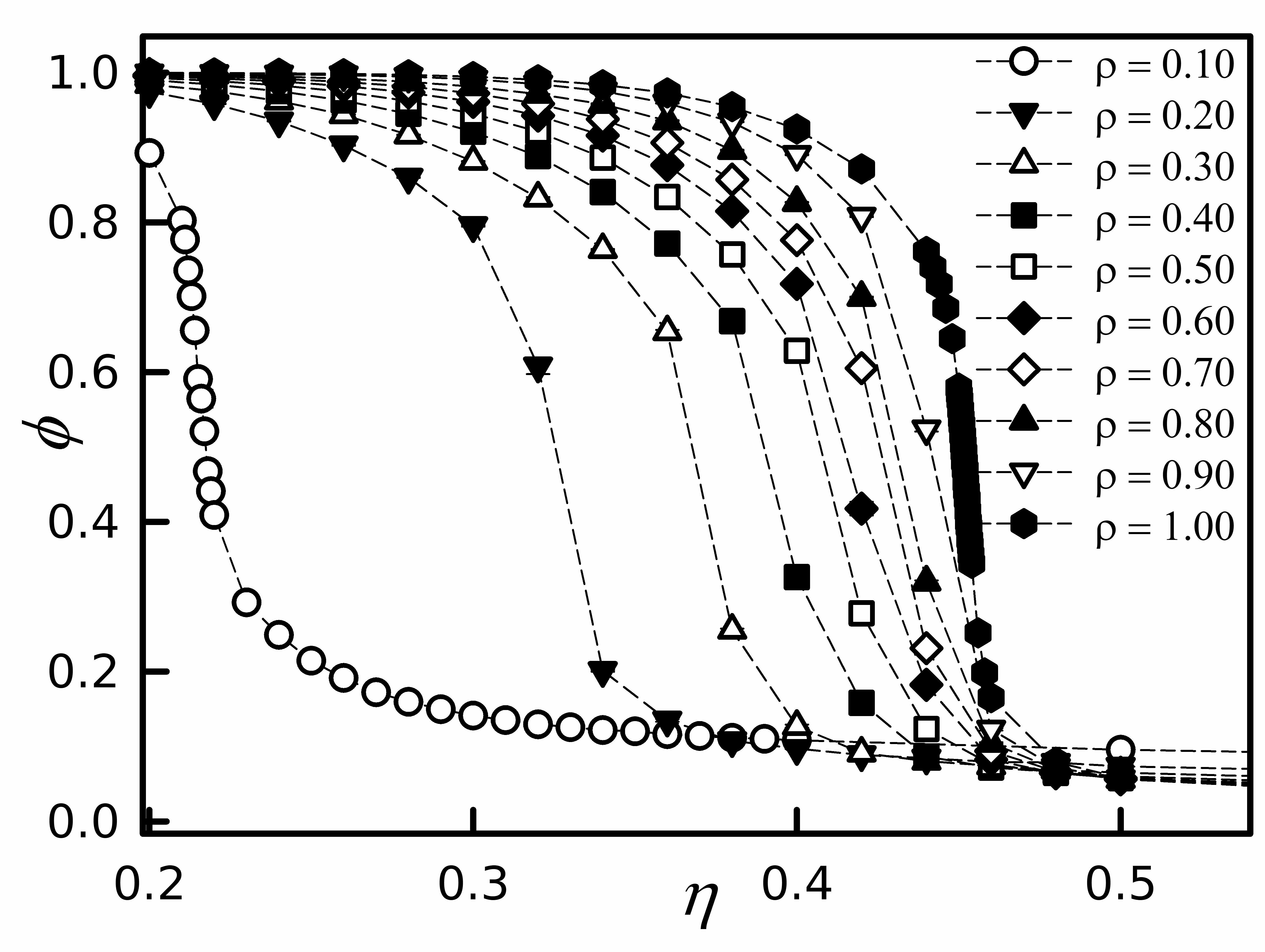}}
    \subfigure[b][] {\includegraphics[width=0.47\linewidth]{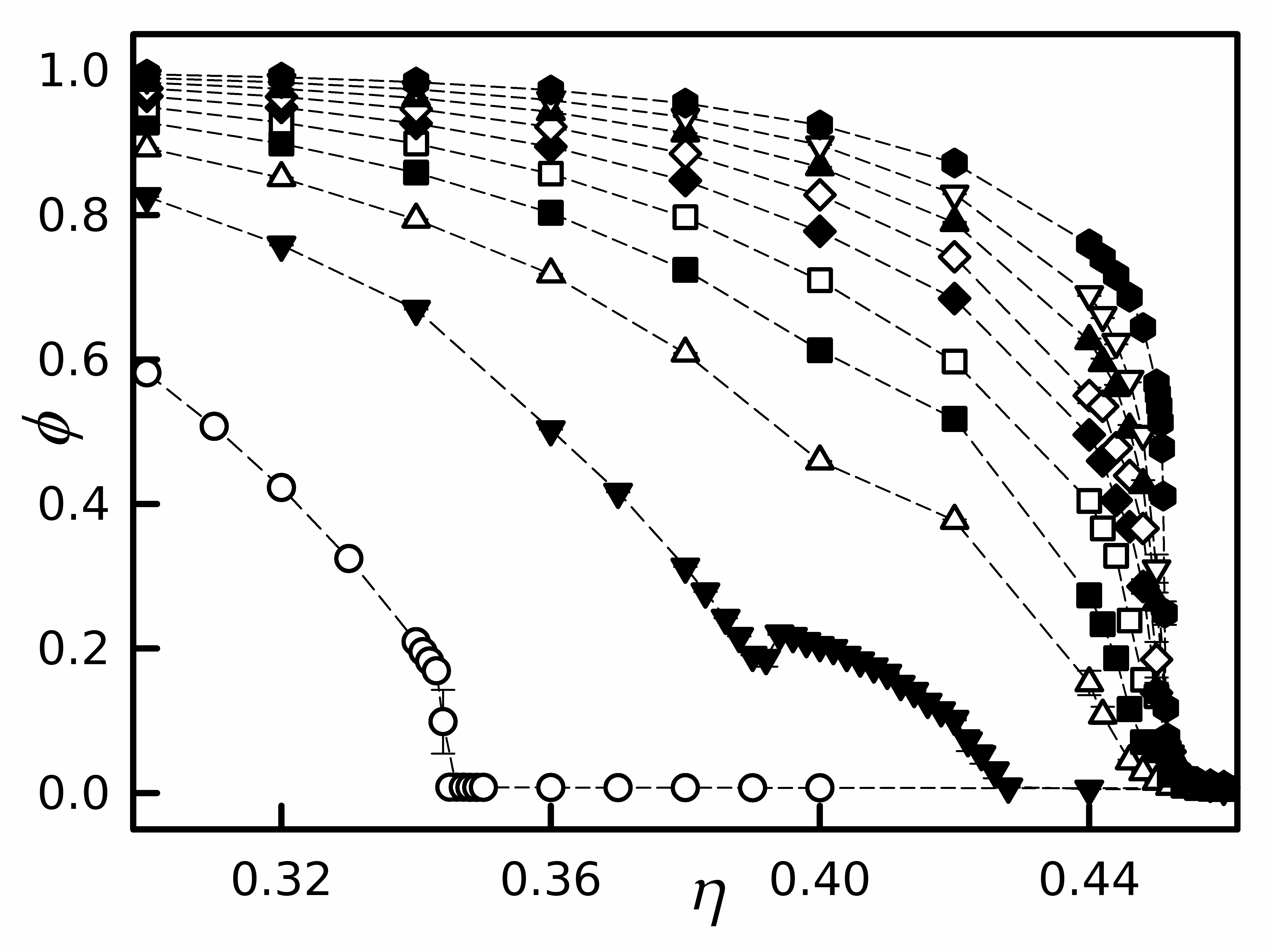}}	
    \caption{\label{fig:phivsetaBI} Order parameter $\phi$ versus $\eta$ for system sizes $L = 33$ (a) and $513$ (b).  Densities as per the legend in (a). In almost all cases, the error bars, which correspond to the standard deviation of the mean, are smaller than the symbols. }  
\end{figure}

In Fig.~\ref{fig:phivsetaBI} the stationary order parameter is plotted versus $\eta$ for $L =33$ and $513$. (For other sizes, see Supplemental Material \cite{supp}.) Comparing these results with those obtained from analysis of configurations (Figs.~\ref{fig:mapConfigurationL513} and \ref{fig:digramL513}),  
one verifies that in the IB phase, $\phi$ decays monotonically with increasing noise. In the DA phase, as expected, $\phi$ tends to zero.

A more complex situation arises when we examine $\phi$ for $\rho = 0.2$ and
$\eta^{-} < \eta < \eta^*$, which includes the intermediate states MBI and TJI
[see Fig.~\ref{fig:phivsetaBI}(b)].
Here, $\phi(\eta)$ is nonmonotonic and exhibits rapid variations in curvature. (Similar 
behavior is observed for $0.2 \leq \rho < 1$.) We do not find a clear correspondence between these variations in $\phi$ and the phase boundaries.  This is not altogether
surprising, since, as noted above, $\phi$ is not a good indicator of traffic-jam states. 

\subsection{\label{nature} Nature of the order-disorder transition}

To identify the nature of the transitions between the disordered and ordered phases, we analyze the Binder cumulant, and look for evidence of finite-size scaling and  hysteresis \cite{tsai1998,furlan2019,chaikin1995}. Here we concentrate on densities $\rho = 0.1$ and $1$ (full occupancy) using IB initial configurations, since there is apparently only one transition (IB/DA) as we vary $\eta$ at these densities.  We defer analysis of the more complex scenarios observed at intermediate densities to future study.

\subsubsection{Full occupancy}

\begin{figure}[!htp]
	\centering
	\includegraphics[width=0.60\linewidth]{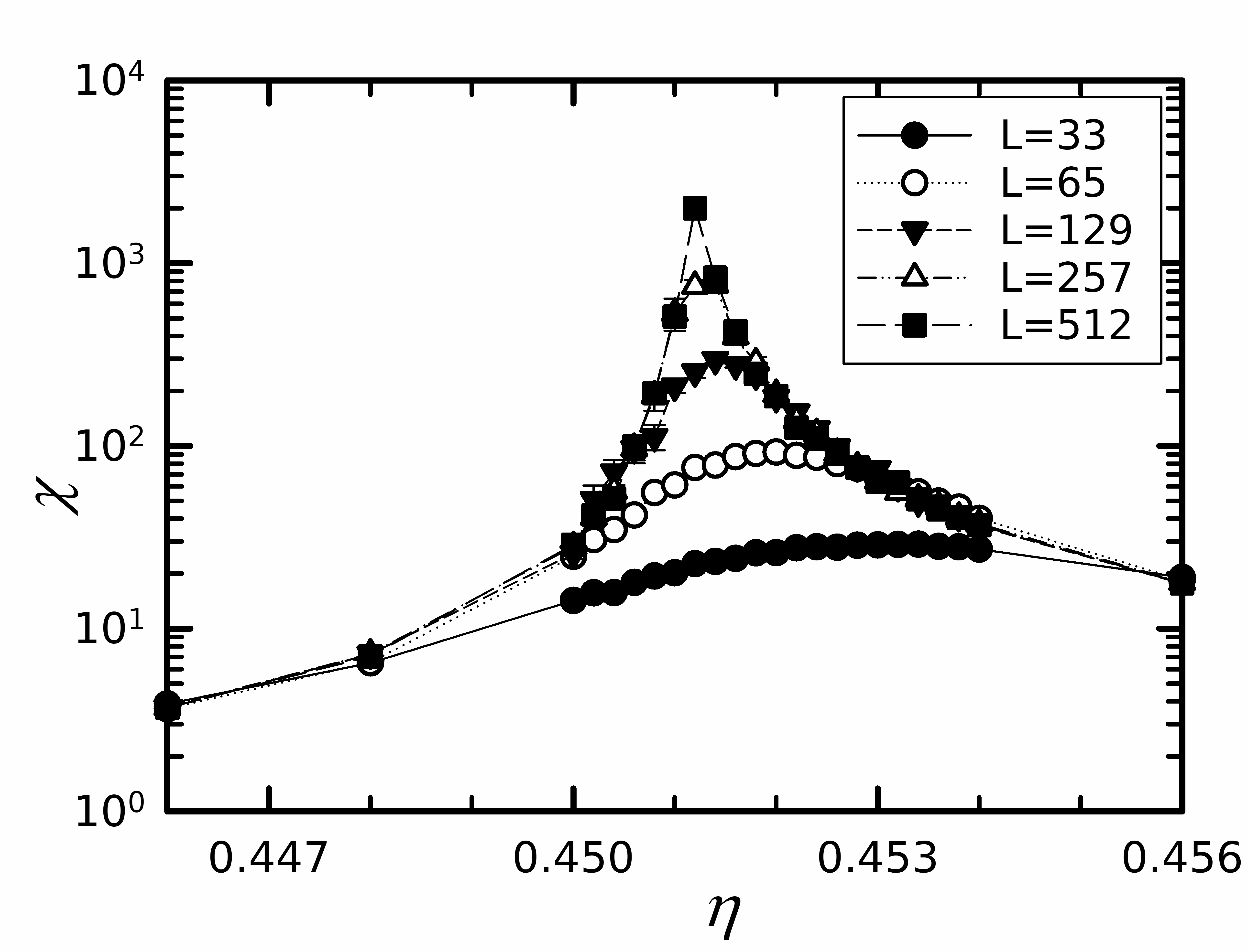}
	\caption{\label{fig:varphi} $\chi$ \textit{vs} $\eta$ for full occupancy, system sizes as indicated.}
\end{figure}

At full occupancy, particles cannot move, so that the model, which enjoys $S3$ (permutation) symmetry, is equivalent to a system of spins, suggesting that the order-disorder transition is continuous, and in the three-state Potts model universality class. The scaled variance, $\chi \equiv L^2 \mbox{var} [\phi]$, exhibits a peak as a function of $\eta$ with an amplitude and sharpness that grows systematically with system size, as expected at a critical point (see Fig.~\ref{fig:varphi}). The position of the maximum approaches a limiting value of $\eta=0.4512(2)$ as the system size tends to infinity.

\begin{figure}[!htp]
    \centering
    \includegraphics[width=0.60\linewidth]{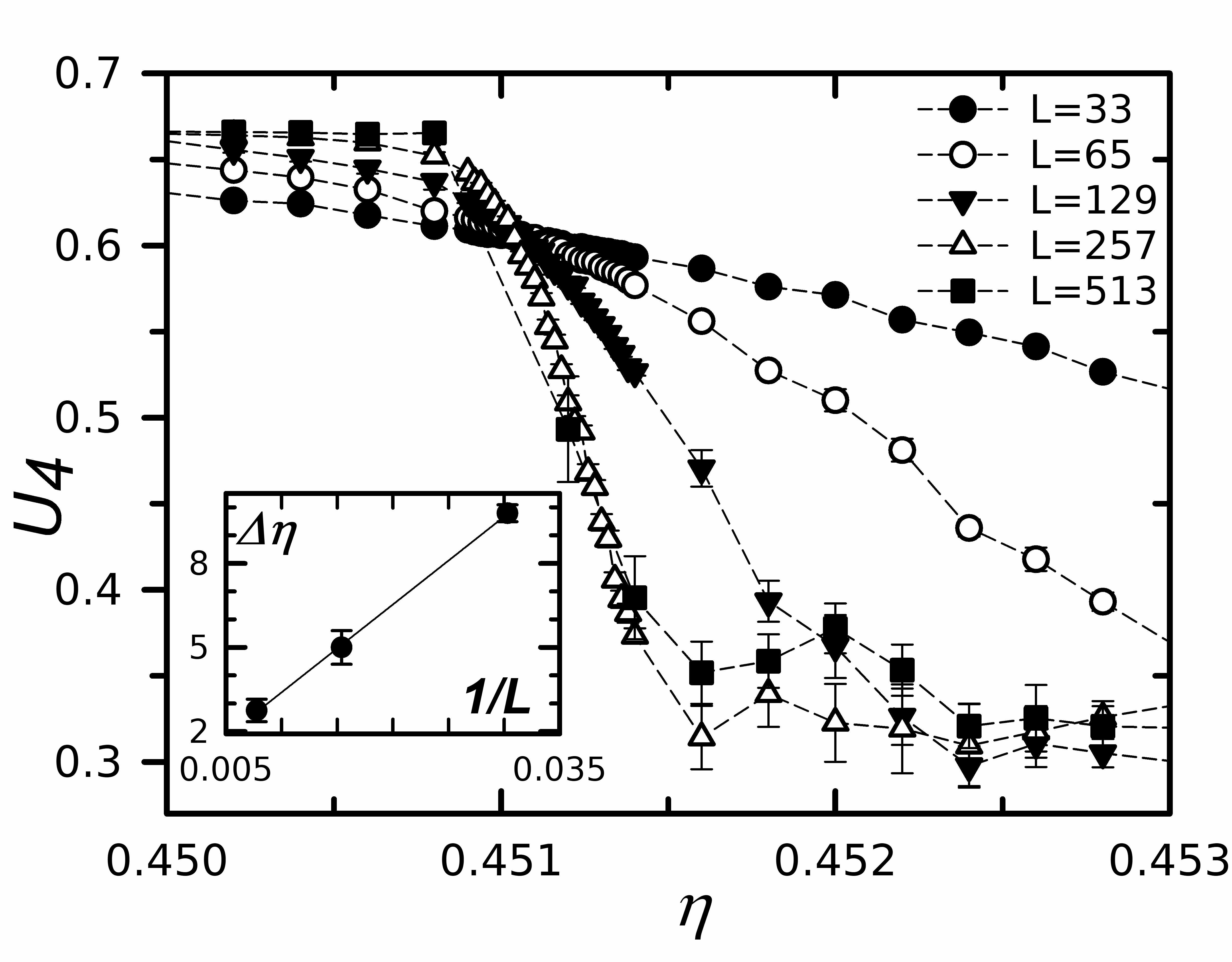}
    \caption{\label{fig:inteceptU4} Binder cumulant $U_4$ versus noise intensity $\eta$ at full occupancy, system sizes as indicated. Inset: $\Delta \eta \equiv (\eta_n-0.45100)\times10^{5}$ (where $\eta_n$ is the value marking the crossing of the $U_4$ curves for sizes $L_n$ and $L_{n+1}$) versus $1/L_{n+1}$. The line is a least-squares linear fit to the data, yielding a limiting value of $\eta = 0.451002(5)$
    (see text).}
\end{figure}

The Binder cumulant \cite{binder1984},

\begin{equation}
	U_4 = 1 - \frac{\langle \phi ^4 \rangle}{3 \langle \phi ^2 \rangle ^2},
\end{equation}

\noindent shown in Fig.~\ref{fig:inteceptU4}, exhibits a series of crossings that converge to a limiting value, $\eta_c$, as expected at a continuous phase transition.

To estimate $\eta_c$ from the cumulant crossings, we perform a series of high-statistics studies for $L= 33$, 65, 129 and 257, using 30-60 independent realizations for each $\eta$ value, and a relaxation time of $2\times10^5$ MC steps followed by production runs of $2.8 \times 10^6$ steps. The inset of Fig.~\ref{fig:inteceptU4} plots $\eta_n$, the value marking the crossing of $U_4$ for sizes $L_n \equiv 2^n +1$ and $L_{n+1}$, versus $1/L_{n+1}$. A least-squares linear fit to these data yields the estimate $\lim_{n \to \infty} \eta_n = 0.451002(6)$.

This estimate for the critical noise intensity is in accord with that obtained via analysis of $\chi$, but considerably more precise.  It also compares well with the result of a mean-field (MF) analysis that treats all 19 sites in the neighborhood of a given particle as statistically independent, yielding $\eta_c = 0.4704$.  (A discrepancy of about $4\%$ between simulation and a simple MF analysis is of course hardly surprising.) Details of this and other MF analyses, including stability estimates for condensed structures such as immobile bands, will be presented in a forthcoming publication \cite{diaz0000}.

Considering the results for the scaled variance and Binder cumulant reported above, as well as the apparent absence of hysteresis in the order parameter, 
we conclude that the phase transition at full occupancy is continuous. This motivates a finite-size scaling analysis, based on the usual scaling hypothesis that at $\eta_c$, the order parameter and its variance follow power laws,
that is, 

\begin{equation}
    \phi \propto L^{-\beta/\nu}, \;\;\;\;\;\; \mbox{and} \;\;\;\;\;\;
    \chi \propto L^{\gamma/\nu},
\end{equation}

\noindent where $\nu$ is the critical exponent governing the growth of the correlation length in the critical region. (While these asymptotic power laws may receive subdominant correction terms for smaller sizes, the present data are insufficient to reliably introduce correction-to-finite-size-scaling terms.)  For off-critical values of $\eta$, plots of $\ln \phi$ or $\ln \chi$ versus $\ln L$ exhibit significant curvature. 

\begin{figure}[!htp]
	\center
	\subfigure[a][] {\includegraphics[width=0.49\linewidth]{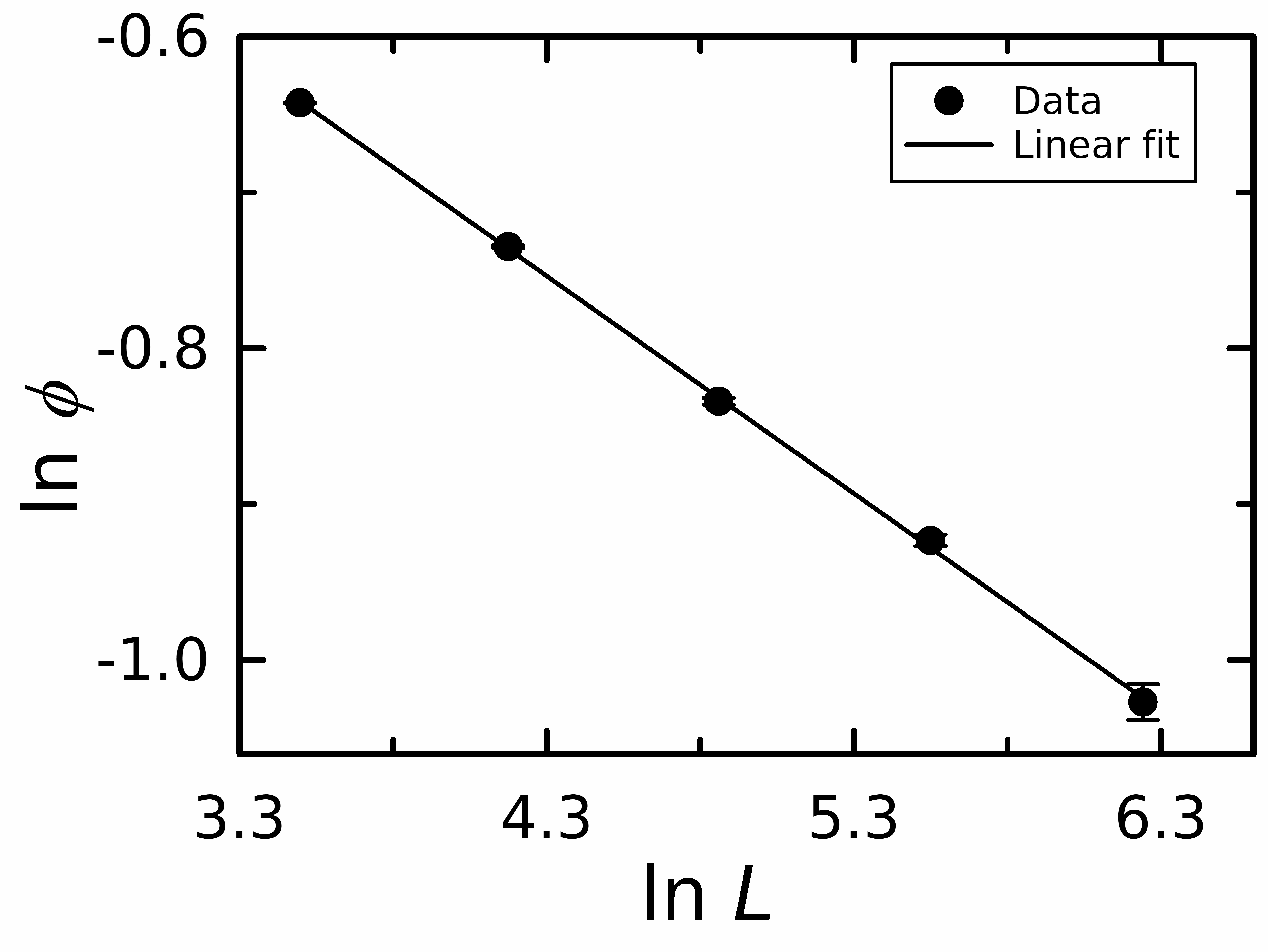}}
	\subfigure[b][] {\includegraphics[width=0.49\linewidth]{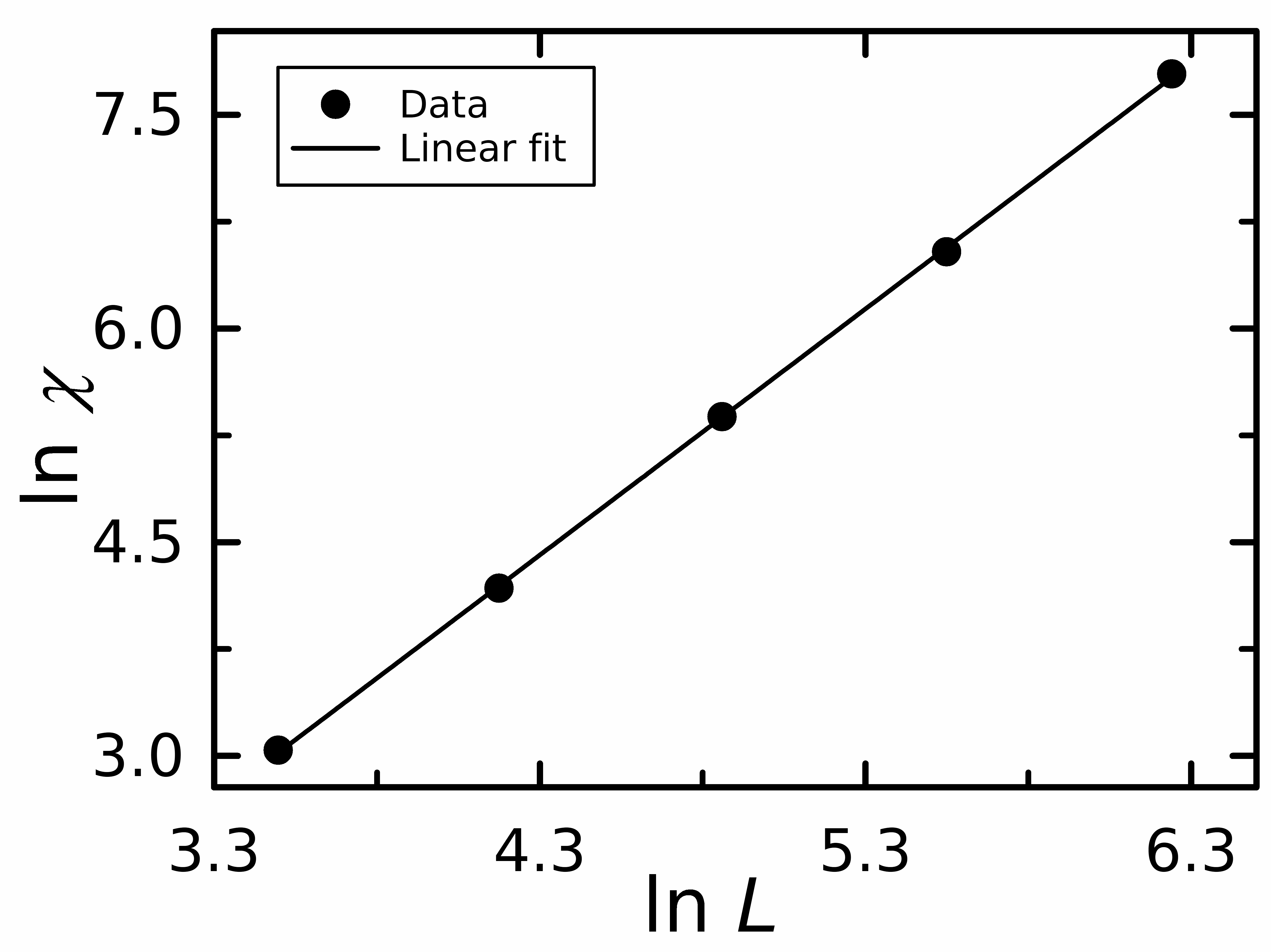}}	
	\caption{\label{fig:finSizeScale} Log-log plots of (a) $\phi$ and (b) $\chi$ for $\eta = 0.451034$. Error bars are smaller than the symbols.}
\end{figure}

Applying this criterion to a set of high-statistics studies, using 30-60 independent realizations (with $10^6$ MC relaxation steps followed by time-averages over the subsequent $2 \times 10^6$ steps, for a series of $\eta$ values in the critical region) we obtain $\eta_c = 0.451034(1)$ and $0.451021(6)$ based on the data for $\phi$ and $\chi$, respectively. Figure \ref{fig:finSizeScale} shows the order parameter and its scaled variance for $\eta=0.451034$ for the five system sizes analyzed in this work; these data are well fit by a linear expression. 

Analysis of the resulting curvatures of log-log plots of the order parameter and its scaled variance versus $L$, combined with the value of $\eta_c$ obtained via analysis of the cumulant crossings, permits us to restrict $\eta_c$ to the interval [0.451030,0.451036]. Interpolating the slopes of the least-squares linear fits to the data for $\ln \phi$ (and $\ln \chi$) as a function of $\ln L$, for $\eta = 0.451030, 0.451032, 0.451034$ and $0.451036$, and taking into account error propagation, we obtain the critical exponent estimates,

\begin{equation}
	\frac{\beta}{\nu} = 0.138(3)\;\;\;\;\text{and}\;\;\;\;\frac{\gamma}{\nu} = 1.70(2).
\end{equation}

The values of the corresponding quantities for the three-state Potts model in two dimensions are $\beta/\nu=2/15=0.133\ldots$ and $\gamma/\nu=26/ 15=1.733\ldots$ \cite{wu1982}.  Thus our estimates for both $\beta/\nu$ and $\gamma/\nu$ are quite close to the Potts model values. The small discrepancies (about $4\%$ and $2\%$ for $\beta/\nu$ and $\gamma/\nu$, respectively) are likely due to the limited range of system sizes and possible corrections to FSS.  (This is reflected in the fact that, while our three estimates for $\eta_c$ differ by less than $0.007\%$, the values, with their respective uncertainties, are mutually incompatible.)

We conclude that at full occupancy, the critical behavior of our model is consistent with that of the three-state Potts model in two dimensions.  This is of course to be expected, since the symmetry of the three-state majority-vote model is precisely that of the corresponding Potts model \cite{zubillaga2022}.

\subsubsection{Density 0.1}

While the phase transition at full occupancy is clearly continuous, our analysis at the much lower density of $0.1$ yields strong evidence of a discontinuous transition between the IB and DA phases. In preliminary studies, we perform simulations extending to $10^6$ MC steps, using two kinds of ICs: (1) a single IB with vapor, and (2) completely random positions and velocities. ICs for case (1) are prepared by allowing thirty independent realizations to relax for $10^6$ MC steps using $\eta = 0.31$ and an IC similar to Fig.~\ref{fig:confinit}(b), and saving the final configuration of each realization. Subsequently, for each independent realization of the ordering studies, we randomly select one of these thirty configurations as the IC and allow the system to relax for an additional $4 \times 10^5$ MC steps at the new value of $\eta$ before collecting data. 

\begin{figure}[!htp]
	\center
	\subfigure[a][] {\includegraphics[width=0.40\linewidth]{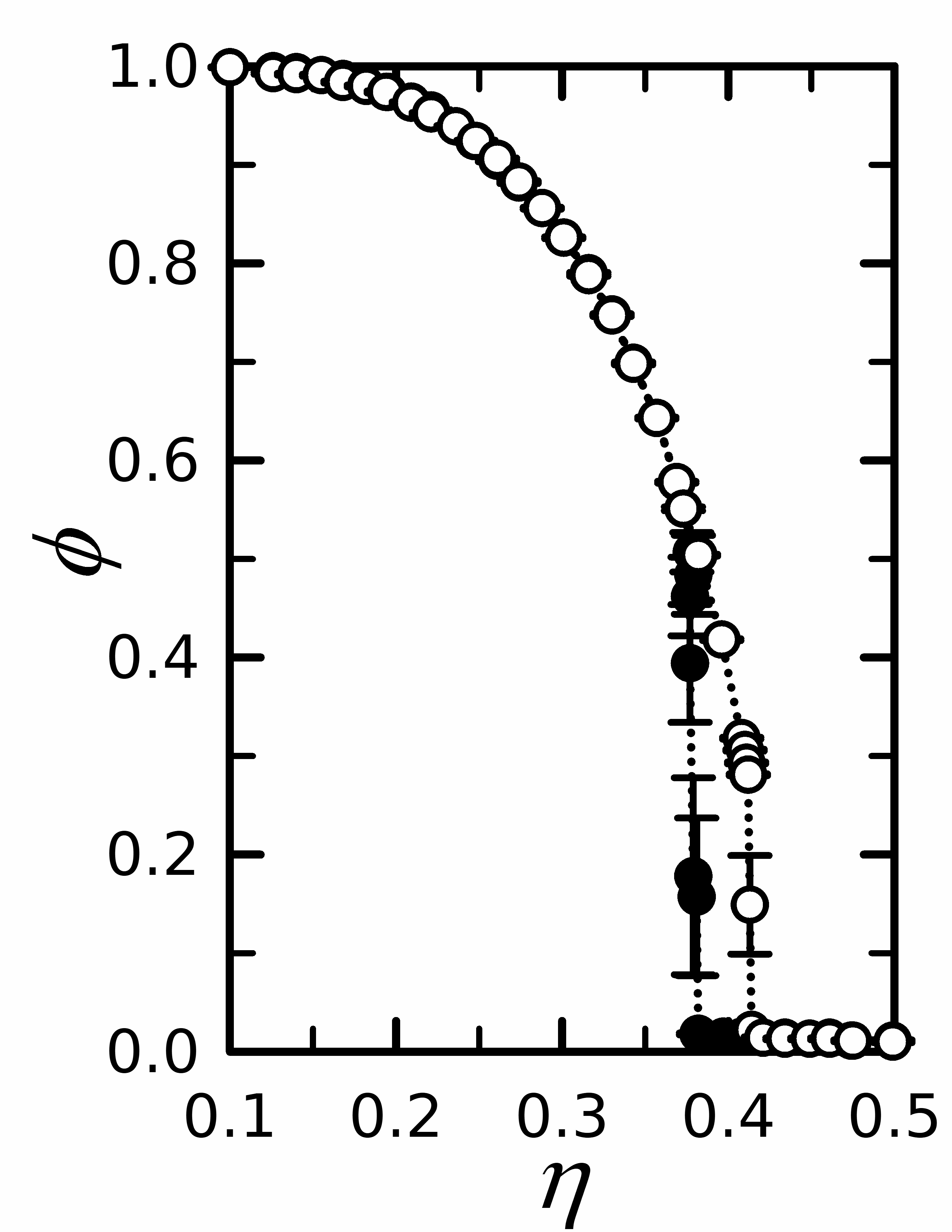}}
	\subfigure[b][] {\includegraphics[width=0.40\linewidth]{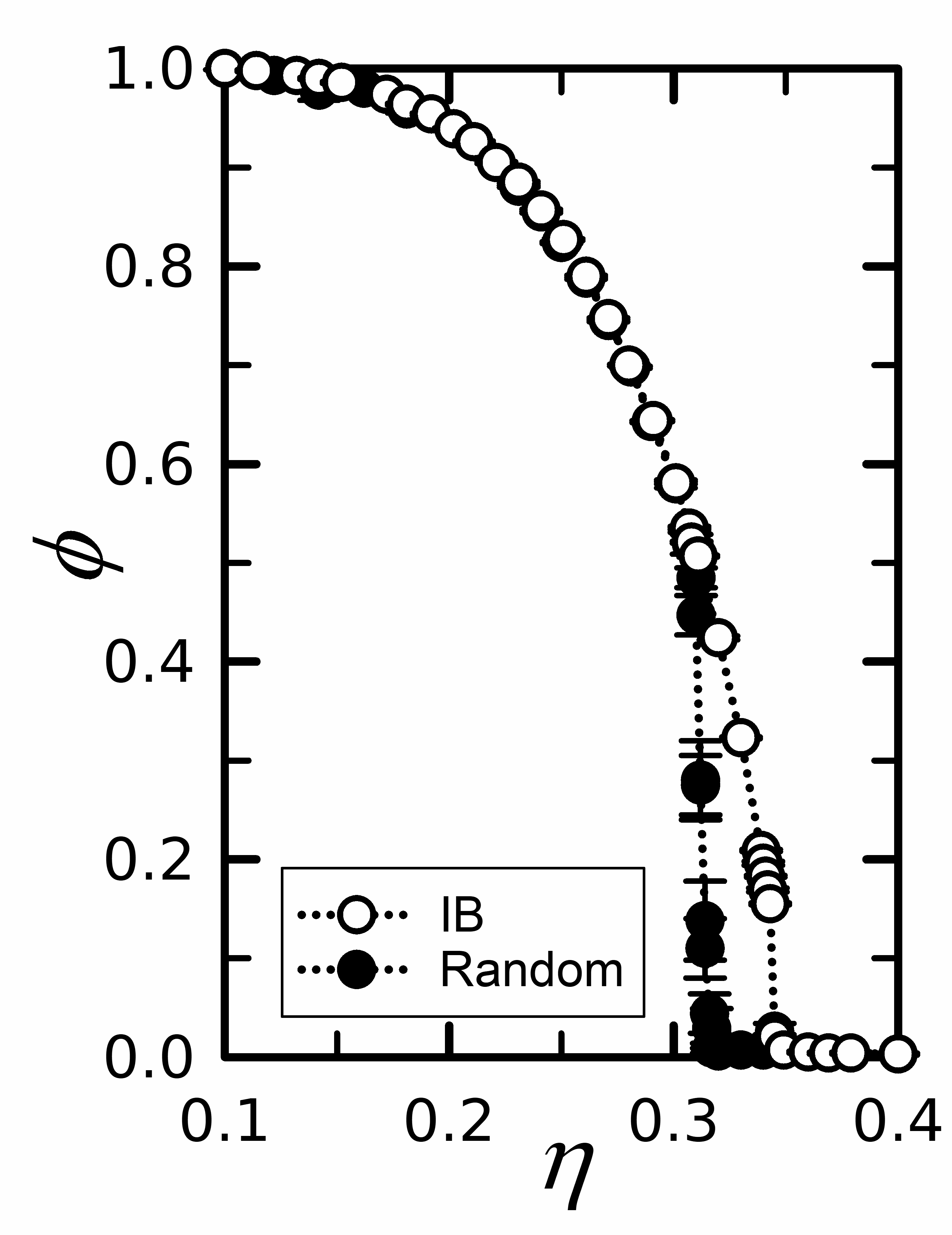}}	
	\caption{\label{fig:hysteresis} Density $\rho = 0.1$. Order parameter $\phi$ vs $\eta$ for an IB initial configuration (empty circle) and for a Random IC (filled circle). System sizes: (a) $L = 257$, (b) $L = 513$.}
\end{figure}

Figure \ref{fig:hysteresis} shows $\phi$ versus $\eta$ for system sizes $L=257$ and $L=513$, yielding several key observations. First, IB is the more stable configuration for low noise values. Second, the steady state depends on the initial configuration for $\eta \in \left[0.310,\ 0.345\right]$. Third, the order parameter appears to jump from a positive value to zero at the IB $\to$ DA transition.

For both of the sizes studied, the curves for $\phi$ obtained using IB and random ICs are essentially identical for $\eta$ in the interval $[0.1, 0.3]$.  In this range, the steady-state configuration remains a single IB for IB initial configurations, and consists of one or two IBs when using random ICs.

\begin{figure}[!htp]
    \centering
    \includegraphics[width=0.60\linewidth]{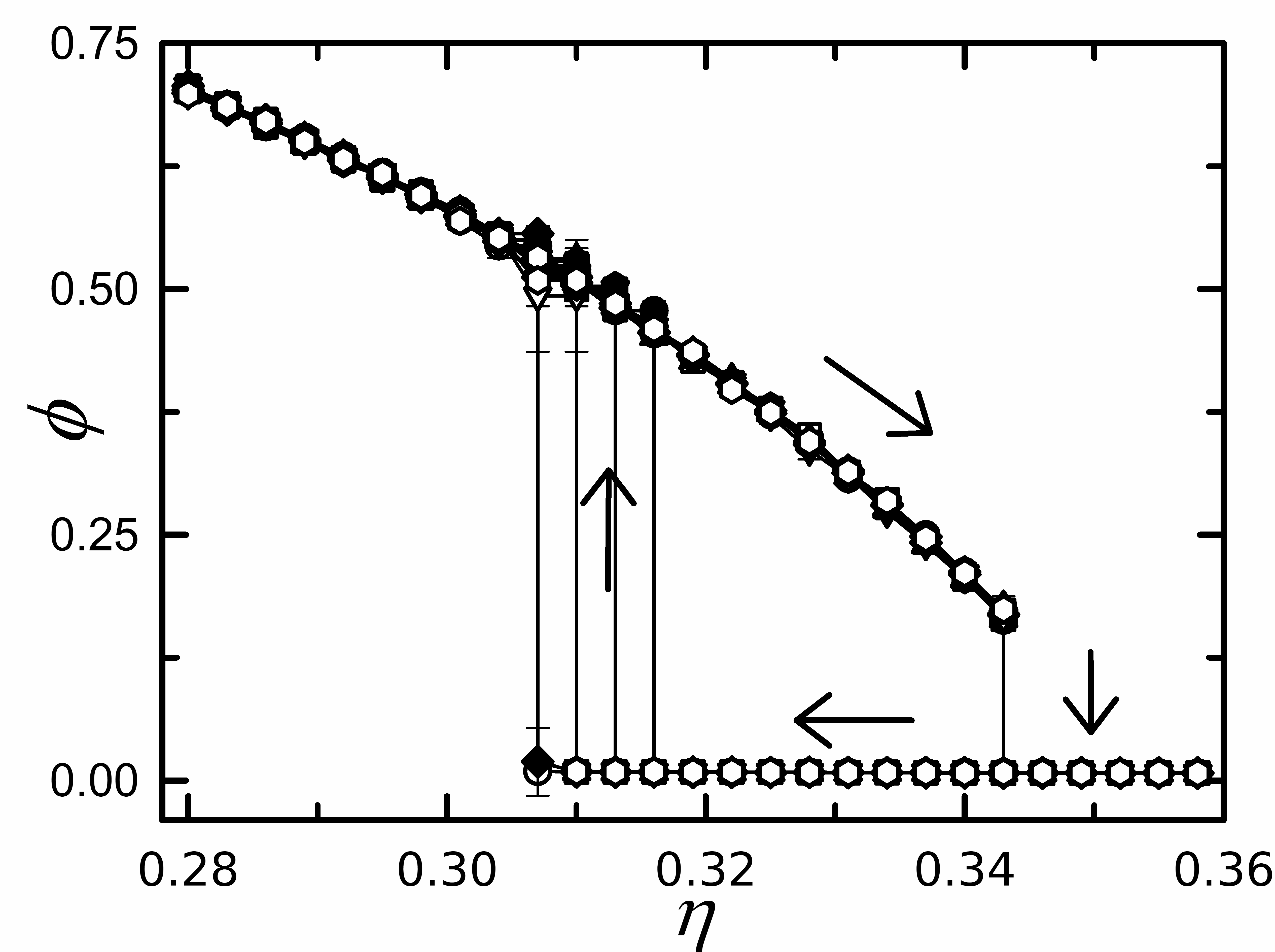}
    \caption{\label{fig:hystL513} Hysteresis curves obtained from twelve independent realizations. Arrows indicate increasing or decreasing $\eta$ and resulting upward or downward jumps in $\phi$. Density 0.1, system size $L=513$. }
\end{figure}

The above results confirm the stability of the IB steady state for low noise, bistability of the steady state for intermediate noise, and an apparent jump in the order parameter at the IB $\to$ DA transition.  Taken together, they motivate a search for hysteresis, which we perform as follows. We generate twelve independent realizations with $L=513$ and $\rho = 0.1$, starting from $\eta = 0.28$, and gradually increase the noise intensity (using increments $\Delta \eta = 0.003$) until the system, initially in the IB phase, exhibits a transition to the DA phase. We subsequently reverse the process, slowly decreasing $\eta$ until the system returns to the IB phase. Following each change in $\eta$, we allow the system to relax for $10^5$ MC steps. Figure~\ref{fig:hystL513} shows $\phi$ versus $\eta$ for the twelve independent realizations; each exhibits a hysteresis loop.

We note that the upper terminal points of the hysteresis loops all fall in the $\eta$ interval $[0.343, 0.346]$, marking the stability limit of the IB phase. By contrast, transitions from DA to IB, occur over a broader interval, $[0.307, 0.316]$.  (The mean and standard deviation over the twelve studies are $0.313$ and $0.002$, respectively.) 

Videos of the evolution near the stability limits (see Supplementary Material \cite{supp}) reveal that in both cases, the transition requires the formation of a critical nucleus: of an IB fragment capable of growing until it wraps the system, in the transition from DA to IB, and of a critical ``bubble" within the IB at the inverse transition.

\begin{figure} [!htp]
	\center
	\includegraphics[width=0.60\linewidth]{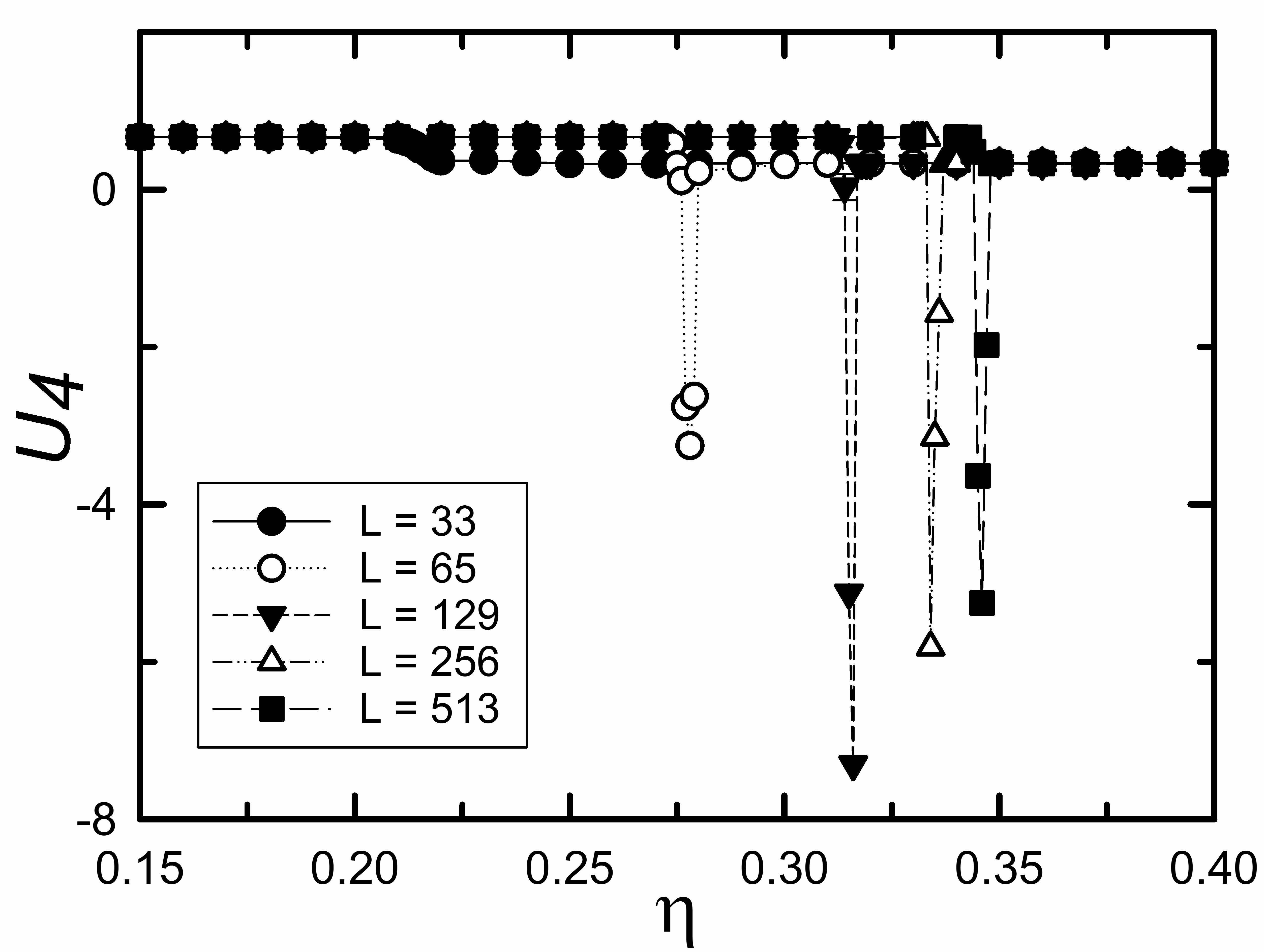}
	\caption{Dependence of $U_4$ upon noise intensity for all sizes and IC IB.}
	\label{fig:lowdensu4}
\end{figure}

We also find clear signs of a discontinuous IB/DA transition in the Binder cumulant. A key signature of a discontinuous phase transition (associated with a bimodal probability distribution for the order parameter - see Supplementary Material \cite{supp}) is nonmonotonic dependence of $U_4$ on noise intensity, including negative values in the vicinity of the transition \cite{tsai1998}. Figure~\ref{fig:lowdensu4} shows $U_4$ for density $\rho = 0.1$. For $L \geq 65$, regardless of the initial configuration, the cumulant assumes negative values in the regions with abrupt changes in $\phi$, indicating that the phase transition is discontinuous. (For $L = 33$ the behavior is typical of a continuous transition, as might be expected for such a small system size.)

In summary, there are strong indications that for density $\rho = 0.1$, the order-disorder transition is discontinuous.  We expect this to hold for even smaller densities, and for some range of densities {\it greater than} $0.1$, although we defer a precise determination of this range, and analysis of scaling at the associated tricritical point, to future study.

\section{\label{sec:concl} Summary and conclusions}

We investigate a Vicseklike model with excluded volume on a triangular lattice, in which particles can only assume three different velocities. The model exhibits a wide variety of steady-state configurations, depending on the noise intensity, density, initial configuration and system size.
Different from Vicseklike models without excluded volume \cite{vicsek1995, chate2008a}, in which the transition to order involves bands of mobile particles, here, excluded volume leads to condensed structures in which most particles are immobile.  For densities smaller than unity, condensed phases coexist with a low-density disordered vapor, whose properties evolve continuously into those of the uniform disordered phase. Of the five condensed phases depicted in Fig.~\ref{fig:typesteadystates}, that consisting of a single IB is the most stable at low noise intensities, regardless of system size, density, or initial configuration. In certain cases, more than one type of configuration can appear for the same set of parameters.

A previous study of an active four-state Potts model by Peruani and coworkers \cite{peruani2011}, revealed ordered phases corresponding (in our terminology) to an immobile band (IB) or a traffic jam (TJI). The present model exhibits these phases as well as two additional condensed structures: a mobile band and a second type of traffic jam. The glider configurations reported in \cite{peruani2011} are not observed here. As noted in Sec.~II, the differences in the phase diagrams are quite plausible, given the differences in lattice structure, number of states, and velocity-update procedure between the two models.  Precisely which differences allow certain states to appear in our model but not in that of Ref.~\cite{peruani2011} nevertheless remains an open question.

Several studies of active matter models - both with and without excluded volume interactions - exhibit a discontinuous transition between ordered and disordered phases \cite{chate2008a, ginelli2016, cambui2016, katyal2020, mangeat2020, romenskyy2013}. 
Similar to the active Potts models studied in \cite{peruani2011,karmakar2022}, the nature of the order-disorder transition in the present model changes from discontinuous, at low density, to continuous at high density. This is similar to the behavior observed by \cite{paoluzzi2024}, in an active-matter model with geometric frustration and a tendency to alignment, in continuous space. While our model exhibits a single phase transition at very high, as well as very low densities, at intermediate densities, we observe transitions between various phases, such as mobile bands and traffic jams, at intermediate noise intensities. At full occupancy, the model can be interpreted as a three-state majority-vote model having the same symmetry as the three-state Potts model.  A finite-size-scaling analysis yields critical exponent ratios of $\beta/\nu = 0.138(3)$ and $\gamma/\nu = 1.70(2)$, in fair agreement with the exact values for the three-state Potts model in two dimensions. 

Our study leaves several open questions for future study: probing the occurrence and stability of the non-IB condensed phases; pinpointing the change from a discontinuous to a continuous order-disorder phase transition; investigating the associated tricritical scaling; and understanding dynamic aspects of the phase transition. 

A refined analysis of the phase boundaries of non-IB condensed phases should include particle mobility and current, which change significantly at the transition. Peruani et al.~\cite{peruani2011} suggested that the change from a continuous to a discontinuous IB/DA transition coincides when the density is that of the site-percolation threshold, placing it at $\rho = 0.5$ on the triangle lattice.  This is certainly possible, although one might question the relevance of independent percolation to a model with nontrivial correlations. In fact, our results for the stability limits $\eta^-$ and $\eta^*$ shown in Fig.~\ref{fig:eta*eta-} reveal a gap for densities as large as 0.96. If these results (for $L=129$) continue to hold for larger system sizes and densities even nearer unity, it would imply that the transition is continuous for {\it all} densities smaller than unity.

While this study has elucidated in some detail the phase behavior of a simple 
active-matter system with EV and a restricted set of velocities, we note that 
hard-core excluded volume interactions with a one particle per site restriction frustrate flocking, since they cause a decoupling of the density and ordering fields. Lattice models that relax the one-particle rule, or that employ soft-core repulsion, are better suited to the study of flocking \cite{chatterjee2020,karmakar2022,mangeat2020}.

Extensions of the present model include (1) incorporating population dynamics, generating correlations between population and order, and (2) including attractive interactions, which should allow one study ordering free from the artifice of periodic boundaries.  

\section*{Acknowledgments}

The authors are grateful to Hugues Chat\'e for helpful discussions and many insightful suggestions.
This work was partially supported by the Brazilian Agencies CAPES and CNPq.


\end{document}